\documentclass{bioinfo}
\usepackage{lipsum}
\usepackage{custom}
\usepackage[ruled,vlined]{algorithm2e}
\usepackage[caption=false]{subfig}
\copyrightyear{2015}
\pubyear{2015}

\newtheorem{theorem}{Theorem}
\newtheorem{lemma}[theorem]{Lemma}
\newtheorem{corollary}{Corollary}[theorem]
\begin{document}
\firstpage{1}

\title[Optimal Seed Solver]{Optimal Seed Solver: Optimizing Seed Selection in Read Mapping}
\author[Hongyi Xin \textit{et~al}]{Hongyi Xin\,$^{1*}$, Richard Zhu\,$^{1}$,
Sunny Nahar\,$^{1}$, John Emmons\,$^{5}$,\\
Gennady Pekhimenko\,$^{1}$, Carl Kingsford\,$^{3}$, Can Alkan\,$^{4*}$ and Onur Mutlu\,$^2$\footnote{to whom correspondence should be addressed}}
\address{$^{1}$Computer Science Department, Carnegie Mellon University, Pittsburgh, PA, 15213, USA.\\
$^{2}$Dept. of Electrical and Computer Engineering, Carnegie Mellon University, Pittsburgh, PA 15213, USA.\\
$^{3}$Computational Biology Department, Carnegie Mellon University, Pittsburgh, PA 15213, USA.\\
$^{4}$Dept. of Computer Engineering, Bilkent University, Bilkent, Ankara, 06800, Turkey.\\
$^{5}$Department of Computer Science and Engineering, Washington University, St. Louis, MO 63130, USA}



\maketitle

\begin{abstract}

\section{Motivation:} Optimizing seed selection is an important problem in read mapping. The number
of non-overlapping seeds a mapper selects determines the sensitivity of the mapper while the total
frequency of all selected seeds determines the speed of the mapper. Modern seed-and-extend mappers
usually select seeds with either an equal and fixed-length scheme or with an inflexible placement
scheme, both of which limit the potential of the mapper to select less frequent seeds to speed up
the mapping process.  Therefore, it is crucial to develop a new algorithm that can adjust both the
individual seed length and the seed placement, as well as derive less frequent seeds.

\section{Results:} We present the \textbf{Optimal Seed Solver} (OSS), a dynamic programming
algorithm that discovers the least frequently-occurring set of $x$ seeds in an $L$-bp read in
$\mathcal{O}(x\times L)$ operations on average and in $\mathcal{O}(x\times L^{2})$ operations in the
worst case. We compared OSS against four state-of-the-art seed selection schemes and observed that
OSS provides a 3-fold reduction of average seed frequency over the best previous seed selection
optimizations.

\section{Availability:}
We provide an implementation of the Optimal Seed Solver in C at:
https://github.com/CMU-SAFARI/Optimal-Seed-Solver

\section{Contact:} \href{hxin@cmu.edu}{hxin@cmu.edu},
				\href{calkan@cs.bilkent.edu.tr}{calkan@cs.bilkent.edu.tr},
				\href{onur@cmu.edu}{onur@cmu.edu}
\end{abstract}

\section{Introduction}

The invention of high-throughput sequencing (HTS) platforms during the past decade triggered a
revolution in the field of genomics. These platforms enable scientists to sequence mammalian-sized
genomes in a matter of days, which have created new opportunities for biological research. For
example, it is now possible to investigate human genome diversity between
populations~\cite{1000GP,1000GP2012}, find genomic variants likely to cause
disease~\cite{Ng2010,Flannick2014}, and study the genomes of ape species~\cite{Marques-Bonet2009,
Scally2012, Ventura2011, Prado-Martinez2013} and ancient hominids~\cite{Green2010, Reich2010,
Meyer2012} to better understand human evolution.

However, these new sequencing platforms drastically increase the computational burden of genome data
analysis. First, billions of short DNA segments (called reads) are aligned to a long reference
genome. Each read is aligned to one or more sites in the reference based on similarity with a
process called \emph{read mapping}~\cite{Flicek2009-mapping}. Reads are matched to locations in the
genome with a certain allowed number of errors: insertions, deletions, and substitutions (which
usually constitute less than 5\% of the read's length).  Matching strings approximately with a
certain number of allowed errors is a difficult problem. As a result, read mapping constitutes a
significant portion of the time spent during the analysis of genomic data.

In seed-and-extend based read mappers such as mrFAST~\cite{Alkan2009}, RazerS3~\cite{razers3},
GEM~\cite{Marco-Sola2012}, SHRiMP~\cite{shrimp} and Hobbes~\cite{hobbes}, reads are partitioned into
several short, non-overlapping segments called \emph{seeds}. Seeds are used as indexes into the
reference genome to reduce the search space and speed up the mapping process. Since a seed is a
subsequence of the read that contains it, every correct mapping for a read in the reference genome
will also be mapped by the seed (assuming no errors in the seed). Mapping locations of the seeds,
therefore, generate a pool of potential mappings of the read. Mapping locations of seeds in the
reference genome are pre-computed and stored in a \emph{seed database} (usually implemented as a
hash table or Burrows-Wheeler-transformation (BWT)~\cite{Burrows94ablock-sorting} with
FM-index~\cite{FM-index}) and can be quickly retrieved through a database lookup.

When there are errors in a read, the read can still be correctly mapped as long as there exists one
seed of the read that is error free. The error-free seed can be obtained by breaking the read into
many non-overlapping seeds; in general, to tolerate $e$ errors, a read is divided into $e+1$ seeds,
and based on the Pigeonhole Principle, at least one seed will be error free.

Potential mapping locations of the seeds are further verified using a weighted edit-distance
calculation (such as Smith-Waterman~\cite{sw} and Needleman-Wunsch~\cite{nw} algorithms) to
examine the quality of the mapping of the read. Locations that pass this final verification step
(i.e., contain fewer than $e$ substitutions, insertions, and deletions) are valid mappings and are
recorded by the mapper for use in later stages of genomic analysis.

Computing the edit-distance is an expensive operation and is the primary computation performed by
most read mappers. In fact, speeding up this computation is the subject of many other works in this
area of research, such as Shifted Hamming Distance~\cite{SHD}, Gene Myers' bit-vector
algorithm~\cite{Myers1999} and SIMD implementations of edit-distance algorithms~\cite{swps3,
Rognes11}. To allow edits, mappers must divide reads into multiple seeds.  Each seed increases the
number of locations that must be verified. Furthermore, to divide a read into more seeds, the
lengths of seeds must be reduced to make space for the increased number of seeds; shorter seeds
occur more frequently in the genome which requires the mapper to verify even more potential
mappings.

Therefore, the key to building a fast yet error tolerant mapper with high sensitivity is to select
many seeds (to provide greater tolerance) while minimizing their frequency of occurrence (or simply
\emph{frequency}) in the genome to ensure fast operation. Our goal, in this work, is to lay a
theoretically-solid groundwork to enable techniques for optimal seed selection in current and
future seed-and-extend mappers.

Selecting the optimal set of non-overlapping seeds (i.e. the least frequent set of seeds) from a
read is difficult primarily because the associated search space (all valid choices of seeds) is
large and it grows exponentially as the number of seeds increases. A seed can be selected at any
position in the read with any length, as long as it does not overlap other seeds. We observe that
there is a significant advantage to selecting seeds with unequal lengths, as possible seeds of equal
lengths can have drastically different levels of frequencies.

Our goal in this paper is to develop an inexpensive algorithm that derives the optimal placement and
length of each seed in a read, such that the overall sum of frequencies of all seeds is minimized.

This paper makes the following contributions:\vspace{-0.25cm}
\begin{itemize}\setlength\itemsep{2pt}
\item Examines the frequency distribution of seeds in the seed database and counts how often seeds
	of different frequency levels are selected using a naive seed selection scheme.
	We confirm the discovery of prior works~\cite{Kielbasa2011} that frequencies are not evenly
	distributed among seeds and frequent seeds are selected more often under a naive seed
	selection scheme. We further show that this phenomenon persists even when using longer
	seeds.

\item Provides an implementation of an optimal seed finding algorithm, \textbf{Optimal Seed Solver},
	which uses dynamic programming to efficiently find the least frequent non-overlapping seeds
	of a given read. We prove that this algorithm always provides the least frequently-occurring
	set of seeds in a read.

\item Gives a comparison of the Optimal Seed Solver and existing seed selection optimizations,
	including Adaptive Seeds Filter in GEM mapper~\cite{Marco-Sola2012}, Cheap K-mer Selection
	in FastHASH~\cite{Xin2013}, Optimal Prefix Selection in Hobbes mapper~\cite{hobbes} and
	spaced seeds in PatternHunter~\cite{patternhunter}. We compare the complexity, memory
	traffic, and average frequency of selected seeds of Optimal Seed Solver with the above four
	state-of-the-art seed selection mechanisms. We show that the Optimal Seed Solver provides
	the least frequent set of seeds among all existing seed selection optimizations at
	reasonable complexity and memory traffic.

\end{itemize}

\section{Motivation}

To build a fast yet error tolerant mapper with high mapping coverage, reads need to be divided into
many, infrequently occurring seeds. In this way, mappers will be able to find all correct mappings
of the read (mappings with small edit-distances) while minimizing the number of edit-distance
calculations that need to be performed. To achieve this goal, we have to overcome two major
challenges: (1) seeds are short, in general, and therefore frequent in the genome; and (2) the
frequencies of seeds vary significantly. Below we provide discussions about each challenge in
greater detail.

Assume a read has a length of $L$ base-pairs (bp) and $x\%$ of it is erroneous (e.g., $L=80$ and
$x\% = 5\%$ implies that are 4 edits). To tolerate $x\% \times L$ errors in the read, we need to
select $x\% \times L + 1$ seeds, which renders a seed to be $L \div (x\% \times L + 1)$-bp long on
average. Given that the desired error rates for many mainstream mappers have been as large as 5\%,
typically the average seed length of a hash-table based mapper is not greater than
16-bp~\cite{Alkan2009, razers3,shrimp,Marco-Sola2012,hobbes}.

\begin{figure}[h]
\centering
\includegraphics[width=0.48\textwidth]{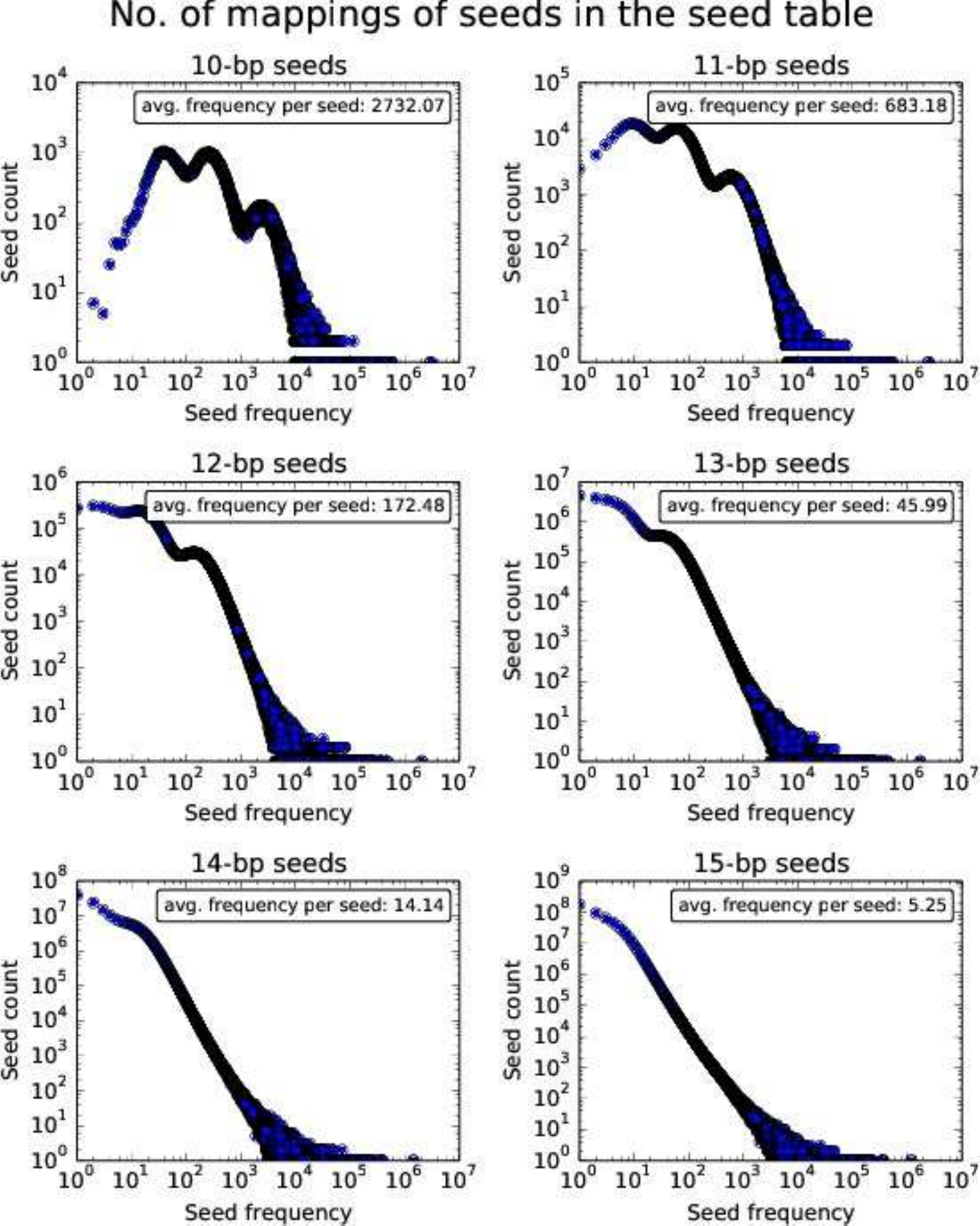}
\caption{Frequency distribution of seeds in fixed-length seed databases. Each plot shows how many
unique seeds there are at each frequency level. Notice that the trend in each plot is not a
continuous line but made of many discrete data points.}
\label{fig:seed_ana}
\end{figure}

Seeds have two important properties: (1) the frequency of a seed is monotonically non-increasing
with greater seed lengths and (2) the frequencies between seeds typically differ (sometimes
significantly)~\cite{Kielbasa2011}. Figure~\ref{fig:seed_ana} shows the static distribution of
frequencies of 10-bp to 15-bp fixed-length seeds from the human reference genome (GRCh37). This
figure shows that the average seed frequency decreases with the increase in the seed length. With
longer seeds, there are more patterns to index the reference genome. Thus each pattern on average is
less frequent.

\begin{figure}[h]
\centering
\includegraphics[width=0.48\textwidth]{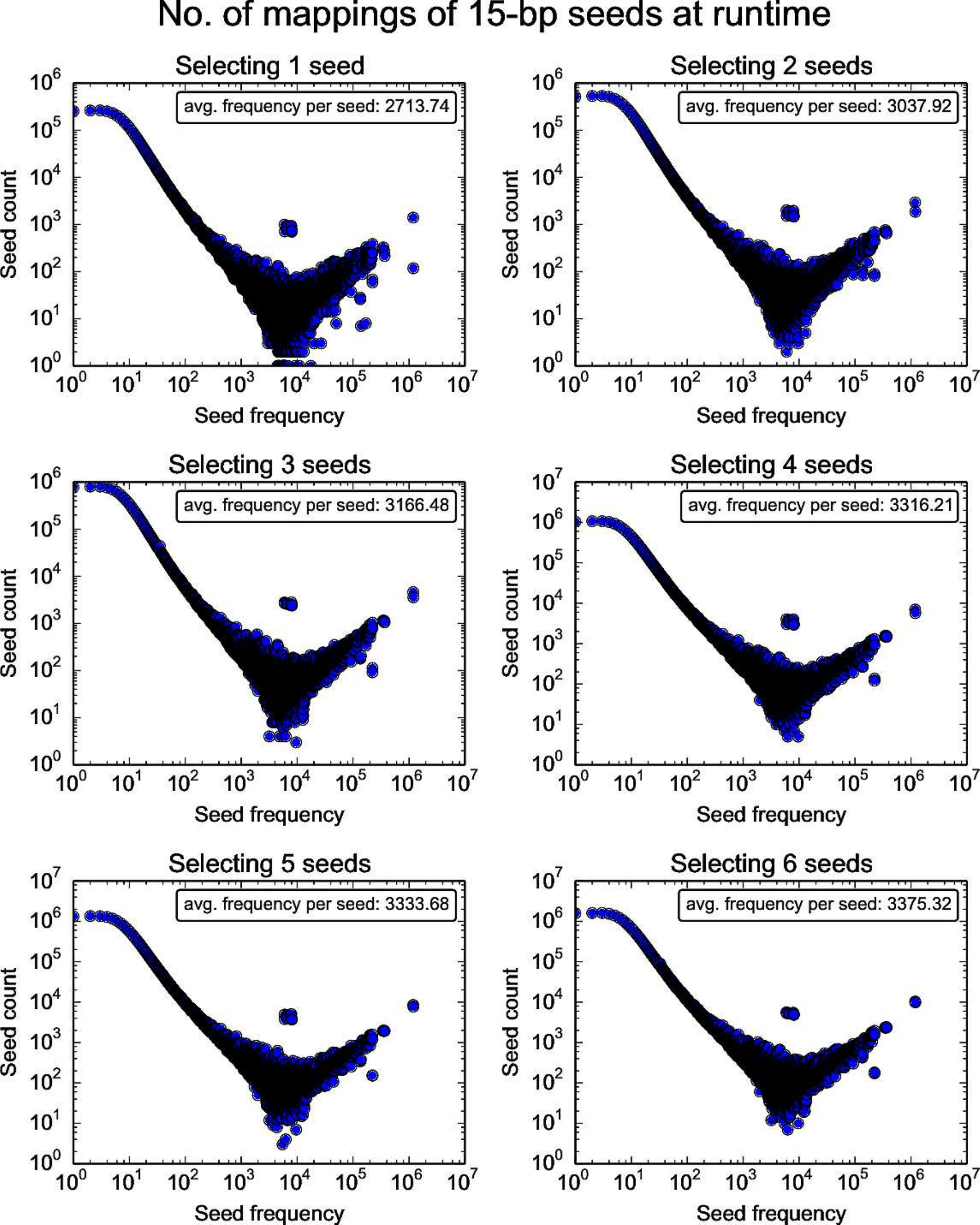}
\caption{Frequency distribution of seeds at runtime by selecting 15-bp seeds consecutively while
mapping 4,031,354 101-bp reads from a real read set, ERR240726 from the 1000 Genome Project, under
different number of required seeds.}
\label{fig:seed_run}
\end{figure}

From Figure~\ref{fig:seed_ana}, we can also observe that the frequencies of seeds are not evenly
distributed: for seeds with lengths between 10 to 15 base-pairs, many seeds have frequencies below
100, as in the figure, seed frequencies below 100 have high seed counts, often over $10^{3}$.
However, there are also a few seeds which have frequencies greater than 100K, even though the seed
counts of such frequencies are very low, usually just $1$. This explains why most plots in
Figure~\ref{fig:seed_ana} follow a bimodal distribution; except for 10-bp seeds and perhaps 11-bp
seeds, where the frequency of seeds peaks at around 100.  Although ultra-frequent seeds (seeds that
appear more frequently than $10^{4}$ times) are few among all seeds, they are ubiquitous in the
genome. As a result, for a randomly selected read, there is a high chance that the read contains one
or more of such frequent seeds.  This effect is best illustrated in Figure~\ref{fig:seed_run}, which
presents the numbers of frequencies of consecutively selected seeds, when we map over 4 million
randomly selected 101-bp reads from the 1000 Genome Project~\cite{1000GP} to the human reference
genome.

Unlike in Figure~\ref{fig:seed_ana}, in which the average frequency of 15-bp seeds is 5.25, the
average frequencies of seeds in Figure~\ref{fig:seed_run} are all greater than 2.7K. Furthermore,
from Figure~\ref{fig:seed_run}, we can observe that the ultra-frequent seeds are selected far more
often than some of the less frequent seeds, as the seed count increases with higher seed frequencies
after $10^{4}$ (as opposed to Figure~\ref{fig:seed_ana}, where seed frequencies over $10^{4}$ usually
have seed counts below $10$). This observation suggests that the ultra-frequent seeds are
surprisingly numerous in reads, especially considering how few ultra-frequent seed patterns there
are in total in the seed database (and the plots in Figure~\ref{fig:seed_run} no longer follow a
bimodal distribution as in Figure~\ref{fig:seed_ana}). We call this phenomenon \emph {frequent
seed phenomenon}. Frequent seed phenomenon is explained in previous works~\cite{Kielbasa2011}. To
summarize, highly frequent seed patterns are ubiquitous in the genome, therefore they appear more
often in randomly sampled reads, such as reads sampled from shotgun sequencing. Frequency distributions of
other seed lengths are provided in the Supplementary Materials.

The key takeaway from Figure~\ref{fig:seed_ana} and Figure~\ref{fig:seed_run} is that although longer
seeds on average are less frequent than shorter seeds, some seeds are still much more frequent than
others and are more prevalent in the reads. Therefore, with a naive seed selection mechanism (e.g.,
selecting seeds consecutively from a read), a mapper still selects many frequent seeds, which
increases the number of calls to the computationally expensive verification process.

To reduce the total frequency of selected seeds, we need an intelligent seed selection mechanism to
avoid using frequent patterns as seeds. More importantly, as there is a limited number of base-pairs
in a read, we need to carefully choose the length of each seed. Extension of an infrequent seed does
not necessarily provide much reduction in the total frequency of all seeds, but it will ``consume''
base-pairs that could have been used to extend other more frequent seeds. Besides determining
individual seed lengths, we should also intelligently select the position of each seed. If multiple
seeds are selected from a small region of the read, as they are closely packed together, seeds are
forced to keep short lengths, which potentially increase their seed frequency.

Based on the above observations, our goal in this paper is to develop an algorithm that can calculate both
the length and the placement of
each seed in the read. Hence, the total frequency of all seeds will be \textbf{minimized}. We call such a
set of seeds the \textbf{optimal seeds} of the read as they produce the minimum number of
potential mappings to be verified while maintaining the sensitivity of the mapper. We call the sum of
frequencies of the optimal seeds the \textbf{optimal frequency} of the read.

\section{Methods}

The biggest challenge in deriving the optimal seeds of a read is the large search space. If we allow a
seed to be selected from an \textbf{arbitrary location} in the read with an \textbf{arbitrary
length}, then from a read of length $L$, there can be $\frac{L\times (L+1)}{2}$ possibilities to
extract a single seed.
When there are multiple seeds, the search space grows exponentially since the position and length
of each newly selected seed depend on positions and lengths of all previously selected seeds. For
$x$ seeds, there can be as many as $\mathcal{O}(\frac{L^{2 \times x}}{x!})$ seed selection schemes.

Below we propose \textbf{Optimal Seed Solver} (OSS), a dynamic programming algorithm that finds
the optimal set of $x$ seeds of a read in $\mathcal{O}(x \times L)$ operations on average and in
$\mathcal{O}(x \times L^{2})$ operations in the worst case scenario.

Although in theory a seed can have any length, in OSS, we assume the length of a seed is bounded by a
range [$S_{min}$, $S_{max}$]. This bound is based on our observation that in practice, neither very
short seeds nor very long seeds are commonly selected in optimal seeds. Ultra-short seeds ($<8$-bp)
are too frequent. Most seeds shorter than 8-bp have frequencies over 1000. Ultra-long seeds
``consume'' too many base-pairs from the read, which shorten the lengths of other seeds and
increase their frequencies. Furthermore, long seeds (e.g., 40-bp) are mostly either
unique or non-existent in the reference genome (seed of 0 frequency is still useful in read
mapping as it confirms there exist at least one error it). Extending a unique or non-existent seed
longer provides little benefit while ``consuming'' extra base-pairs from the read.

Bounding seed lengths reduces the search space of optimal seeds. However, it is not
essential to OSS. OSS can still work without seed length limitations (to lift the limitations, one
can simply set $S_{min} = 1$ and $S_{max} = L$), although OSS will perform extra computation.

Below we describe our Optimal Seed Solver algorithm in three sections. First, we introduce the
\emph{core algorithm} of OSS. Then we improve the algorithm with two optimizations, \emph{optimal
divider cascading} and \emph{early divider termination}. Finally we explain the overall algorithm
and provide the pseudo code.

\subsection{The core algorithm}

A naive brute-force solution to find the optimal seeds of a read would be systematically iterating
through all possible combinations of seeds. We start with selecting the first seed by instantiating all
possible positions and lengths of the seed. On top of each position and length of the first seed, we
instantiate all possible positions and lengths of the second seed that is sampled \textbf{after} (to
the right-hand side of) the first seed. We repeat this process for the rest of the seeds until we
have sampled all seeds. For each combination of seeds, we calculate the total seed frequency and find
the minimum total seed frequency among all combinations.

\begin{figure}[h]
  \centering
  \includegraphics[width=0.48\textwidth]{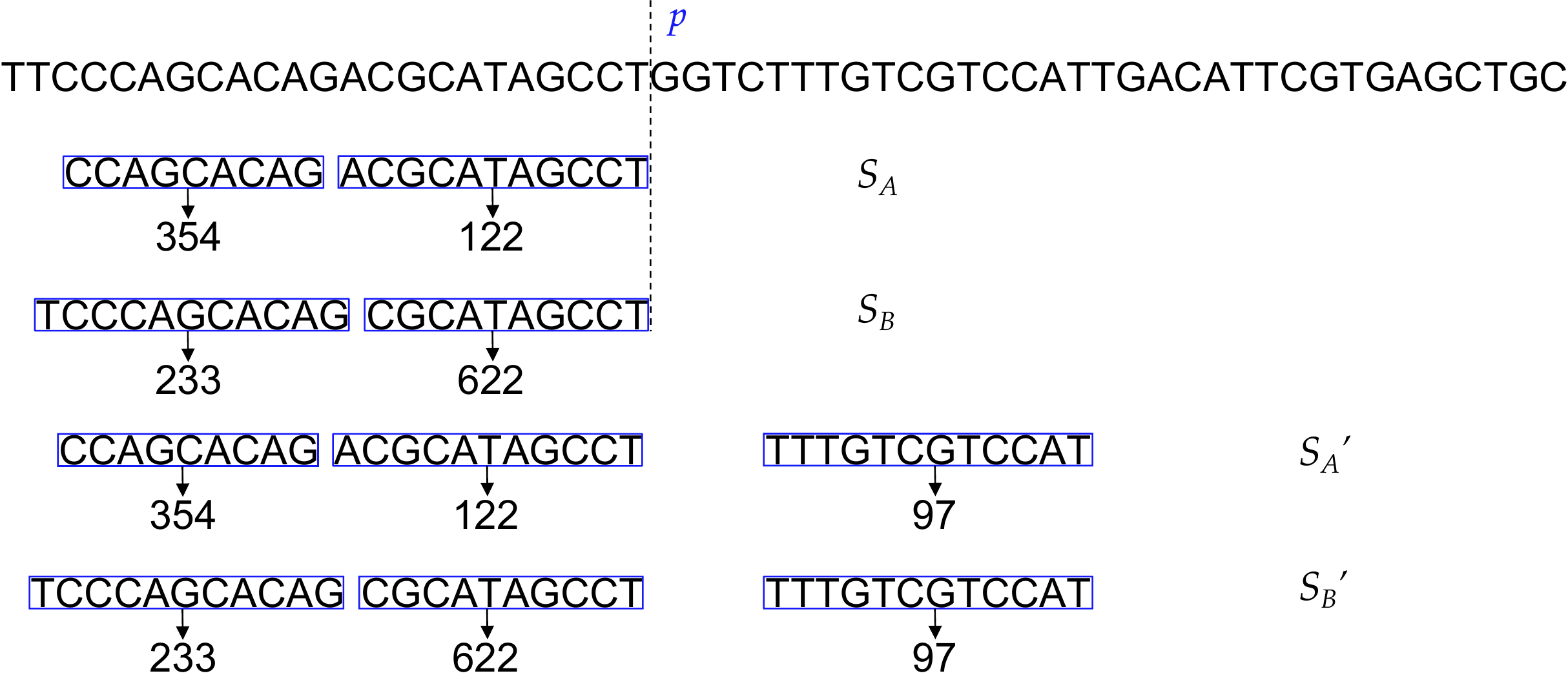}
  \caption{Two 2-seed combinations, $S_{A}$ and $S_{B}$ that end at the same position, $p$, in the read.
  The total seed frequency of $S_{A}$ is smaller than $S_{B}$. Both combinations can be extended by
  adding a third seed, making them $S_{A}'$ and $S_{B}'$ respectively. For any third seed, the total seed
  frequency of $S_{B}'$ must be greater than $S_{A}'$. Hence, $S_{B}$ must not be part of any
  optimal solution.}
  \label{fig:brute-force}
\end{figure}

The key problem in the brute-force solution above is that it examines a lot of obviously suboptimal
combinations. For example, in Figure~\ref{fig:brute-force}, there are two $m$-seed combinations,
$S_{A}$ and $S_{B}$, extracted from the same read, $R$. Both combinations \textbf{end} at the same
position, $p$, in $R$. Among them, $S_{A}$ is more ``optimal'' than $S_{B}$ as it has a smaller total seed
frequency. Then for any number of seeds that is greater than $m$, we know that in the final optimal
solution of $R$, seeds before position $p$ will not be exactly like $S_{B}$, since any seeds that
are appended after $S_{B}$ (e.g., $S_{B}'$ in Figure~\ref{fig:brute-force}) can also be appended
after $S_{A}$ (e.g., $S_{A}'$ in Figure~\ref{fig:brute-force}) and produce a smaller total seed
frequency. In other words, compared to $S_{B}$, only $S_{A}$ has the potential to be part of the
optimal solution and worth appending more seeds after. In general, among two combinations that have
equal numbers of seeds and end at the same position in the read, only the combination with the
smaller total seed frequency has the \textbf{potential} of becoming part of a bigger (more seeds)
optimal solution. Therefore, for a partial read and all combinations of a subset set of seeds in this partial read,
only the optimal subset of seeds of this partial read (with regard to different numbers of seeds)
\textbf{might be} relevant to the optimal solution of the entire read. Any other suboptimal
solutions of this partial read (with regard to different numbers of seeds) will not lead to the
optimal solution and should be pruned.

The above observation suggests that by summarizing the optimal solutions of partial reads under a
smaller number of seeds, we can prune the search space of the optimal solution. Specifically,
given $m$ (with $m < x$) seeds and a substring $U$ that starts from the beginning of the read, only the
optimal $m$ seeds of $U$ \emph{could be} part of the optimal solution of the read. Any other
suboptimal combinations of $m$ seeds of $U$ should be pruned.

Storing the optimal solutions of partial reads under a smaller number of seeds also helps speed up
the computation of larger numbers of seeds. Assuming we have already calculated and stored optimal
the optimal frequency of $m$ seeds of \emph{all} substrings of $R$, to calculate the optimal
$(m+1)$-seed solution of a substrings, we can iterate through a series of divisions of this
substring.
In each division, we divide the substring into two parts: We extract $m$ seeds from the first part
and 1 seed from the second part. The minimum total seed frequency of this division (or simply the
``\emph{optimal frequency of the division}'') is simply the sum of the optimal $m$-seed frequency of
the first part and the optimal 1-seed frequency of the second part. As we already have both the
optimal $m$-seed frequency of the first part and the 1-seed frequency of the second part calculated
and stored, the optimal frequency of this division can be computed with one addition and two
lookups.

The \emph{optimal} $(m+1)$-seed solution of this substring is simply the division that yields the
minimum total frequency. Given that each seed requires at least $S_{min}$ base-pairs, for a
substring of length $L'$, there are in total $L'-(m+1)\times S_{min}$ possible divisions to be
examined. This relationship can be summarized as a recurrence function in
Equation~\ref{eq:recur}, in which $Opt(U, m)$ denotes the optimal $m$-seed frequency of substring
$U$ and $u$ denotes the length of $U$.

\begin{equation}
Opt(U, m+1) = \min_{i} Opt(R[1:i-1], m) + Opt(R[i:u], 1)
\label{eq:recur}
\end{equation}

OSS implements the above strategy using a dynamic programming algorithm: To calculate the optimal
$(x+1)$-seed solution of a read, $R$, OSS computes and stores optimal solutions of partial reads
with fewer seeds through $x$ iterations. In each iteration, OSS computes optimal solutions of
substrings with regard to a specific number of seeds. In the $m^{th}$ iteration ($m \leq x$), OSS
computes the optimal $m$-seed solutions of \textbf{all} substrings that \textbf{starts from the
beginning of} $R$, by re-using optimal solutions computed from the previous $(m-1)^{th}$
iteration. For each substring, OSS performs a series of divisions and finds the division that
provides the minimum total frequency of $m$ seeds. For each division, OSS computes the optimal
$m$-seed frequency by summing up the optimal $(m-1)$-seed frequency of the first part and the
1-seed frequency of the second part. Both frequencies can be obtained from previous iterations.
Overall, OSS starts from 1 seed and iterates to $x$ seeds. Finally OSS computes the optimal
$(x+1)$-seed solution of $R$ by finding the optimal division of $R$ and reuses results from the
$x^{th}$ iteration.

\subsection{Further optimizations}

With the dynamic programming algorithm, OSS can find the optimal $(x+1)$ seeds of a $L$-bp read in
$\mathcal{O}(x\times L^{2})$ operations: In each iteration, OSS examines $\mathcal{O}(L)$ substrings
$(L-i\times S_{min}$ substrings for the $i^{th}$ iteration$)$ and for each substring OSS inspects
$\mathcal{O}(L)$ divisions $(L'-i\times S_{min}$ divisions of a $L'$-bp substring$)$. In total, there
are $\mathcal{O}(L^{2})$ divisions to be verified in an iteration.

To further speed up OSS and reduce the average complexity of processing each iteration, we
propose two optimizations to OSS: optimal divider cascading and early divider termination.
With both optimizations, we empirically reduce the average complexity of processing an iteration to
$\mathcal{O}(L)$.  Below we describe both optimizations in detail.

\subsubsection{Optimal divider cascading}

Until this point, our assumption is  that optimal solutions of substrings within an iteration are independent from
each other: the \emph{optimal division} (the division that provides the optimal frequency of the
substring) of one substring is independent from the optimal division of another substring, thus
they must be derived independently.

We observe that this assumption is not necessarily true as there exists a relationship between
two substrings of different lengths in the same iteration (under the same seed number): the \emph{first
optimal divider} (the optimal divider that is the closest towards the beginning of the read, if
there exist multiple optimal divisions with the same total frequency) of the \emph{shorter}
substring must be \textbf{at the same or a closer position towards the beginning of the read}, compared to
the \emph{first optimal divider} of the \emph{longer} substring. We call this phenomenon the \emph{optimal
divider cascading}, and it is depicted in Figure~\ref{fig:cascade}. The proof that
the optimal divider cascading is always true is provided in the Supplementary Materials.

\begin{figure}[h]
  \centering
  \includegraphics[width=0.48\textwidth]{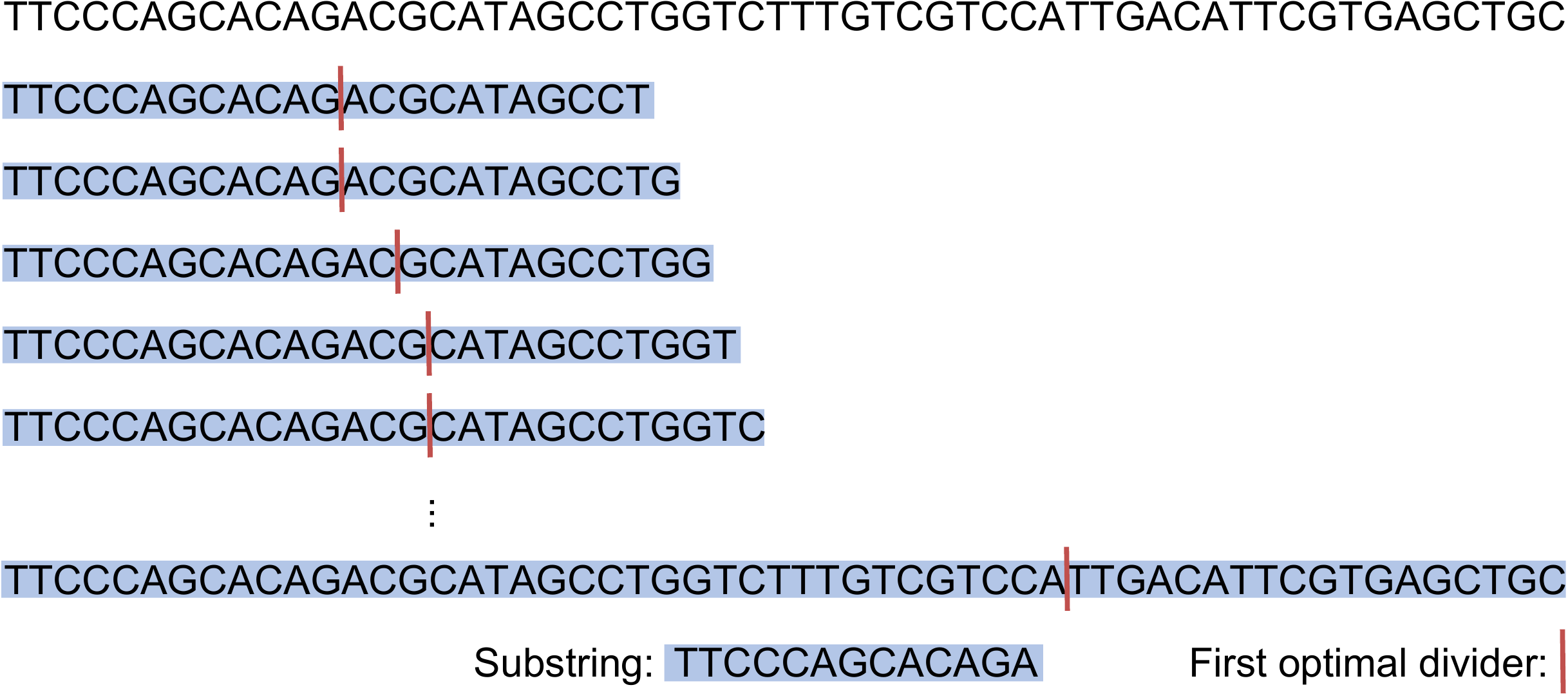}
  \caption{All substrings and their first optimal dividers in an iteration. We observe that the first
  optimal divider of a \textbf{longer} substring is never more towards the beginning of the read
  than the first optimal divider of a \textbf{shorter} substring.}
  \label{fig:cascade}
\end{figure}

Based on the optimal divider cascading phenomenon, we know that for two substrings in the same
iteration, the first optimal divider of the shorter substring must be no further than the first
optimal divider of the longer substring. With this relationship, we can reduce the search space of
optimal dividers in each substring by processing substrings within an iteration from the longest to
the shortest.

In each iteration, we start with the longest substring of the read, which is \textbf{the read
itself}. We examine all divisions of the read and find the \textbf{first optimal divider} of it.
Then, we move to the next substring of the length $|L-1|$. In this substring, we only need to check
dividers that are at the same or a prior position than the first optimal divider of the read. After
processing the length $|L-1|$ substring, we move to the length $|L-2|$ substring, whose search space
is further reduced to positions that are at the same or a closer position to the beginning of the
read than the first optimal divider of the length $|L-1|$ substring. This procedure is repeated
until the shortest substring in this iteration is processed.

\subsubsection{Early divider termination}

With optimal divider cascading, we are able to reduce the search space of the \emph{first} optimal
divider of a substring and exclude positions that come after the first optimal divider of the
previous, 1-bp longer substring. However, the search space is still large since any divider prior to
the first optimal divider of the previous substring could be the optimal divider. To further reduce
the search space of dividers in a substring, we propose the second optimization -- \emph{early divider
termination}.

The key idea of early divider termination is simple: The optimal frequency of a substring
monotonically non-increases as the substring extends longer in the read (see
Lemma~1 in Supplementary Materials for the proof of this fact).

Based on the optimal divider cascading, we start at the position of the first optimal divider in the
previous substring. Then, we gradually move the divider towards the beginning (or simply moving
backward) and check the total seed frequency of the division after each move. During this process,
the first part of the division gradually shrinks while the second part gradually grows, as
we show in Figure~\ref{fig:mov_div}. According to the Lemma~1 in the Supplementary
Materials, the optimal frequency of the first part must be monotonically non-decreasing while the
optimal frequency of the second part must be monotonically non-increasing.

\begin{figure}[h]
  \centering
  \includegraphics[width=0.40\textwidth]{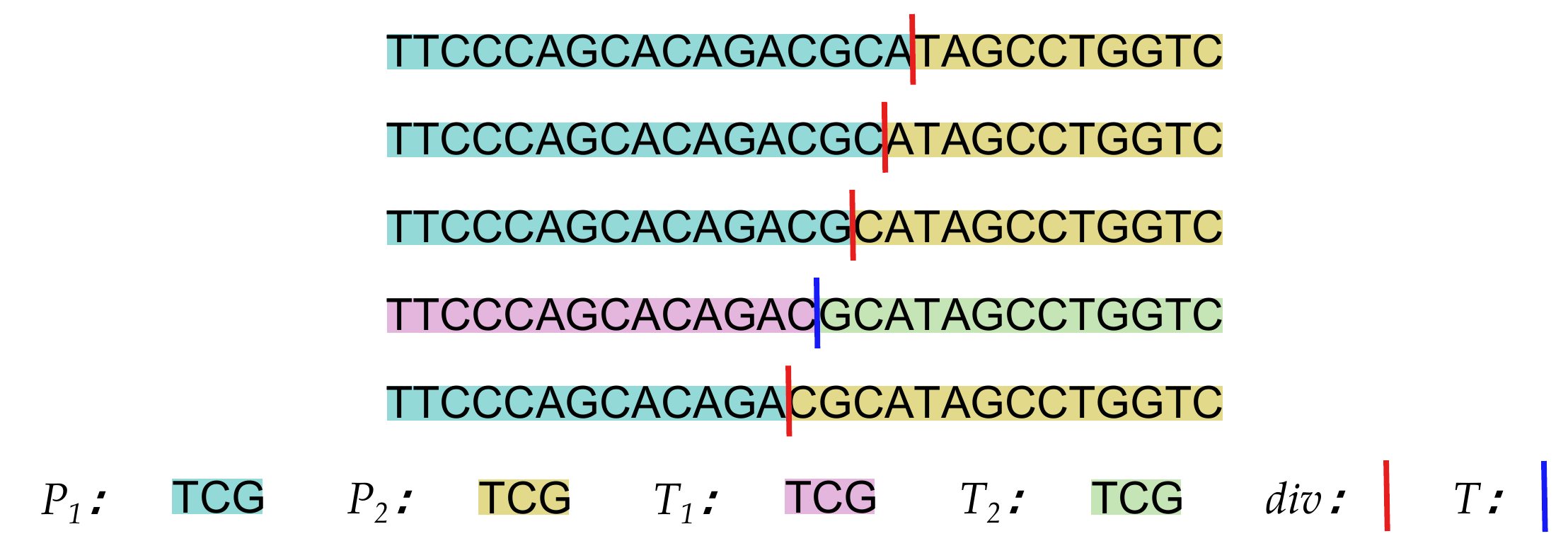}
  \caption{Propagating dividers in a substring. The divider starts at the position of the
  previous substring's first optimal divider, then gradually moves towards the beginning of the
  substring, until it reaches the termination position, $T$.
  }
  \label{fig:mov_div}
\end{figure}

For each position of the divider, let $\mathtt{FREQ}_{P_2}$ denote the frequency of the \emph{second part}
$(P_2)$ and $\Delta \mathtt{FREQ}_{P_1}$ denote the change of frequency of the \emph{first part} $(P_1)$
between current and the next move. Early divider termination suggests that: the divider should stop
moving backward, whenever $|\Delta \mathtt{FREQ}_{P_1}| > |\mathtt{FREQ}_{P_2}|$. All dividers that are prior to this
position are guaranteed to have greater total seed frequencies. We call this stopping position
the \emph{termination position}, and the division at this position -- \emph{termination division}, denoted
as $T$. We name the first and the second part of $T$ as $T_1$ and $T_2$ respectively.

For any divider that comes prior to the termination position, compared to the termination division, its
first part is shorter and its second part is longer. Hence the optimal frequency of its first part
is greater and the optimal frequency of its second part is smaller. Let $|\Delta \mathtt{FREQ}_{P_1'}(T)|$
denote the increase of the optimal frequency of the first part between current division and termination
division and $|\Delta \mathtt{FREQ}_{P_2'}(T)|$ denote the decrease of the second part. Based on
Lemma~1, we have $|\Delta \mathtt{FREQ}_{P_1'} (T)| \geq |\Delta \mathtt{FREQ}_{T_1}|$. Since the
frequency of a seed can be no smaller than 0, we also have $|\mathtt{FREQ}_{T_2}| \geq |\Delta
\mathtt{FREQ}_{P_2'}(T)|$. Combining the three inequalities, we have $|\Delta \mathtt{FREQ}_{P_1}(T)| > |\Delta
\mathtt{FREQ}_{P_1}(T)|$. This suggests that compared to the termination division, the frequency increase of
the first part must be greater than the frequency reduction of the second part. Hence, the overall
optimal frequency of such division must be greater than the optimal frequency of the termination
division. Therefore divisions prior to termination position cannot be optimal.

Using early divider termination, we can further reduce the search space of dividers within a
substring and exclude all positions that are prior to the termination position. Since the second
part of the substring hosts only one seed and frequencies of most seeds decrease to 1 after
extending it to a length of over 20 bp, we observe that the termination position of a substring is reached
fairly quickly, only after a few moves. With both optimal divider cascading and early divider
termination, from our experiments, we observe that we only need to verify 5.4 divisions on average
for each substring. To conclude, with both optimizations, we have reduced the average complexity of
Optimal Seed Solver to $\mathcal{O}(x \times L)$.

\subsection{The full algorithm}

Below we present the full algorithm of Optimal Seed Solver. Before calculating the optimal $x$-seed
frequency of the read, $R$, we assume that we already have the optimal 1-seed frequency of any
substring of $R$ and it can be retrieved in a $\mathcal{O}(1)$-time lookup via the
$optimalFreq(\mathrm{substring})$ function. This information can be pre-processed prior to mapping,
or it can be calculated dynamically at runtime. If calculated at runtime, it requires at most
$\mathcal{O}(L^{2})$ lookups to the seed database for all possible substrings of the read.


Let $firstOptDivider(\mathrm{substring})$ be the function to calculate the first
optimal divider of a substring. Then the optimal set of seeds can be calculated by filling a
2-D array, $opt\_data$, of size $(x-1)\times L$. In this array, each element stores two data: an
optimal seed frequency and a first optimal divider. For the element at $i^{th}$ row and $j^{th}$
column, it stores the optimal $i$-seed frequency of the substring $R[1...j]$ as well as the first
optimal divider of the substring.

\begin{algorithm}[t]
  \textbf{Input}: the read, $\mathsf{R}$\\
  \textbf{Output}: the optimal $x$-seed frequency of $\mathsf{R}$, $\mathsf{opt\_freq}$ and the first $x$-seed optimal divider of R, $\mathsf{opt\_div}$\\
  \textbf{Global data structure}: the 2-D data array $\mathsf{opt\_data[\ ][\ ]}$\\
  \textbf{Functions}:\\
  $\mathit{firstOptDivider}$: computes the first optimal divider of the substring\\
  $\mathit{optimalFreq}$: retrieves the optimal 1-seed frequency of a substring\\
  \textbf{Pseudocode}:\\
  \tcp{Fill the first iteration}
  \For{$\mathit{l} = L$ \KwTo $S_{min}$} {
	  $\mathsf{substring} = \mathsf{R[\mathit{1...l}]}$\;
	  $\mathsf{opt\_data[\mathit{1}][\mathit{l}].freq} = \mathit{optimalFreq(\mathsf{substring})}$\;
  }
  \tcp{From iteration 2 to x-1}
  \For{$\mathit{iter} = 2$ \KwTo $x-1$} {
	  \tcp{Previous optimal divider}
	  $\mathsf{prev\_div} = L - S_{min} + 1$\;
	  \For{$\mathit{l} = L$ \KwTo $\mathit{iter}\times S_{min}$} {
		  $\mathsf{substring} = \mathsf{R[1\mathit{...l}]}$\;
		  $\mathsf{div} = \mathit{firstOptDivider(\mathsf{substring}, \mathit{iter}, \mathsf{prev\_div})}$\;
		  $\mathsf{1st\_part} = \mathsf{R[1\mathit{...div}-1]}$\;
		  $\mathsf{2nd\_part} = \mathsf{R[\mathit{div...L}]}$\;
		  $\mathsf{1st\_freq} = \mathsf{opt\_data[\mathit{iter}-1][\mathit{div}-1].freq}$\;
		  $\mathsf{2nd\_freq} = \mathit{optimalFreq(\mathsf{2nd\_part})}$\;
		  $\mathsf{opt\_data[\mathit{iter}][\mathit{l}]}.div = \mathsf{div}$\;
		  $\mathsf{opt\_data[\mathit{iter}][\mathit{l}]}.freq = \mathsf{1st\_freq} + \mathsf{2nd\_freq}$\;
		  \tcp{optimal seed cascading}
		  $\mathsf{prev\_div} = \mathsf{div}$\;
	  }
  }
  $\mathsf{prev\_div} = L - S_{min} + 1$\;
  $\mathsf{opt\_div} = \mathit{firstOptDivider(\mathsf{R}, L - S_{min} + 1)}$\;
  $\mathsf{1st\_part} = \mathsf{R[1\mathit{...opt\_div}-1]}$\;
  $\mathsf{2nd\_part} = \mathsf{R[\mathit{opt\_div...L}]}$\;
  $\mathsf{1st\_freq} = \mathsf{opt\_data[\mathit{x}-1][\mathit{opt\_div}-1].freq}$\;
  $\mathsf{2nd\_freq} = \mathit{optimalFreq(\mathsf{2nd\_part})}$\;
  $\mathsf{opt\_freq} = \mathsf{1st\_freq} + \mathsf{2nd\_freq}$\;

  \KwRet{$\mathsf{opt\_freq}$, $\mathsf{opt\_div}$}\;
  \caption{optimalSeedSolver}
  \label{algo:A1}
\end{algorithm}

\begin{algorithm}[t]
  \textbf{Output}: the first optimal divider of $\mathsf{substring}$, $\mathsf{opt\_div}$ \\
  \textbf{Global data structure}: the 2-D data array $\mathsf{opt\_data[\ ][\ ]}$\\
  \textbf{Functions}:\\
  $\mathit{optimalFreq}$: retrieves the optimal 1-seed frequency of a substring\\
  \textbf{Pseudocode}:\\
  $\mathsf{first\_div} = \mathsf{prev\_div}$\;
  $\mathsf{min\_freq} = \mathsf{MAX\_INT}$\;
  $\mathsf{prev\_1st\_freq} = \mathsf{MAX\_INT}$\;
  $\mathsf{prev\_2nd\_freq} = \mathsf{MAX\_INT}$\;
  \For{$\mathit{div} = \mathsf{prev\_div}$ \KwTo $(\mathit{iter}-1)\times S_{min}$} {
		  $\mathsf{1st\_part} = \mathsf{substring[1\mathit{...div}-1]}$\;
		  $\mathsf{2nd\_part} = \mathsf{substring[\mathit{div...end}]}$\;
		  $\mathsf{1st\_freq} = \mathsf{opt\_data[\mathit{iter}][\mathit{div}-1].freq}$\;
		  $\mathsf{2nd\_freq} = \mathit{optimalFreq(\mathsf{2nd\_part})}$\;
		  \tcp{early divider termination}
		  \If{$(\mathsf{1st\_freq} - \mathsf{prev\_1st\_freq}) > \mathsf{prev\_2nd\_freq}$} {
			  $\mathit{break}$\;
		  }
		  $\mathsf{freq} = \mathsf{1st\_freq} + \mathsf{2nd\_freq}$\;
		  \tcp{update the optimal divider}
		  \If{$(\mathsf{freq} \leq \mathsf{min\_freq}$} {
			  $\mathsf{min\_freq} = \mathsf{freq}$\;
			  $\mathsf{first\_div} = \mathsf{div}$\;
		  }
		  $\mathsf{prev\_1st\_freq} = \mathsf{1st\_freq}$\;
		  $\mathsf{prev\_2nd\_freq} = \mathsf{2nd\_freq}$\;
  }
  \KwRet{$\mathsf{first\_div}$}\;
  \caption{firstOptDivider}
  \label{algo:A2}
\end{algorithm}

Algorithm~\ref{algo:A1} provides the pseudo-code of $optimalSeedsFinder$ and Algorithm~\ref{algo:A2}
provides the pseudo code of $firstOptDivider$.

To retrieve the starting and ending positions of each optimal seed, we can backtrack the 2-D array
and backward induce the optimal substrings and their optimal dividers in each iteration. The pseudo
code of the backtracking process is provided in Supplementary Materials.

\section{Related Works}

The primary contribution of this work is a dynamic programming algorithm that derives the optimal
non-overlapping seeds of a read in $\mathcal{O}(x \times L)$ operations on average. To our knowledge, this is
the first work that calculates the optimal seeds and the optimal frequency of a read. The most
related prior works are optimizations to the seed selection mechanism which reduce the sum of seed
frequencies of a read using greedy algorithms.

Existing seed selection optimizations can be classified into three categories: (1) extending seed
length, (2) avoiding frequent seeds and (3) rebalancing frequencies among seeds. Optimizations in the
first category extend frequent seeds longer in order to reduce their frequencies. Optimizations in
the second category sample seed positions in the read and reject positions that generate frequent
seeds. Optimizations in the third category rebalance frequencies among seeds such that the average
seed frequency at runtime is more consistent with the static average seed frequency of the seed
database.

We qualitatively compare the Optimal Seed Solver (OSS) to four representative prior
works selected from the above three categories. They are: \emph{Cheap K-mer Selection (CKS)} in
FastHASH~\cite{Xin2013}, \emph{Optimal Prefix Selection (OPS)} in Hobbes~\cite{hobbes},
\emph{Adaptive Seeds Filter (ASF)} in GEM mapper~\cite{Marco-Sola2012} and \emph{spaced seeds} in
PatternHunter~\cite{patternhunter}.  Among the four prior works, ASF represents works from the first
category; CKS and OPS represent works from the second category and spaced seeds represents works
from the third category. Below we elaborate each of them in greater details.

The \textbf{Adaptive Seeds Filter (ASF)} seeks to reduce the frequency of seeds by extending the lengths of the seeds.
For a read, ASF starts the first seed at the very beginning of the read and keeps extending
the seed until the seed frequency is below a pre-determined threshold, $t$. For each subsequent seed, ASF starts it from
where the previous seed left off in the read, and repeats the extension process until the last seed is found.
In this way, ASF guarantees that all seeds have a frequency below $t$.

Compared to OSS, ASF has two major drawbacks. First, it does not allow any
flexibility in seed placement. Seeds are always selected consecutively, starting from the beginning
of the read. Second, it sets a fixed frequency threshold $t$ for all reads.

Selecting seeds consecutively starting at the beginning of a read does not always produce infrequent
seeds. Although most seeds that are longer than 20-bp are either unique or non-existent in the
reference, there are a few seeds that are still more frequent than 100 occurrences even at 40-bp (e.g., all
``A''s). With a small $S_{max}$ (e.g., $S_{max} \leq 40$) and a small $t$ ($t \leq 50$), ASF cannot
not guarantee that all selected seeds are less frequent than $t$. This is because ASF cannot extend
a seed more than $S_{max}$-bp, even if its frequency is still greater than $t$. If a seed starts at
a position that yields a long and frequent seed, ASF will extend the seed to $S_{max}$ and accept a
seed frequency that is still greater than $t$.

Setting a static $t$ for all reads further worsens the problem. Reads are drastically different. Some
reads do not include any frequent short patterns (e.g., 10-bp patterns) while other reads have one
to many highly frequent short patterns. Reads without frequent short patterns do not produce
frequent seeds in ASF, unless $t$ is set to be very large (e.g., $\geq 10,000$) and as a result the
selected seeds are very short (e.g., $\leq 8$-bp). Reads with many frequent short patterns have a
high possibility of producing longer seeds under medium-sized or small $t$'s (e.g., $\leq 100$). For a batch of
reads, if the global $t$ is set to a small number, reads with many frequent short patterns will have
a high chance of producing many long seeds that the read does not have enough length to support. If
$t$ is set to a large number, reads without any frequent short patterns will produce many short but
still frequent seeds as ASF will stop extending a seed as soon as it is less frequent than $t$, even
though the read affords longer and less frequent seeds.

\begin{table*}[pt!]
\center
\begin{tabular}{ ccccccc }
	& Optimal Seed Solver & ASF & CKS & OPS & Spaced seeds & naive\\
	\hline
	Empirical average case complexity & $\mathcal{O}(x\times L)$ & $\mathcal{O}(x)$ & $\mathcal{O}(x\times log \frac{L}{k})$ & $\mathcal{O}(x\times
	L)$& $\mathcal{O}(x)$ & $\mathcal{O}(x)$\\
	\hline
	Number of lookups & $\mathcal{O}(L^{2})$ & $\mathcal{O}(x)$ & $\mathcal{O}(\frac{L}{k})$ & $\mathcal{O}(L)$ & $\mathcal{O}(x)$ & $\mathcal{O}(x)$ \\
\end{tabular}
\caption{An average case complexity and memory traffic comparison (measured by the number
of seed-frequency lookups) of seed selection optimizations, including Optimal Seed Solver (OSS),
Adaptive Seeds Filter (ASF), Cheap K-mer Selection (CKS), Optimal Prefix Selection (OPS), spaced
seeds and naive (selecting fixed-length seeds consecutively). Note that only OSS has different
empirical average case complexity and worst case complexity. The average case and worst case
complexity of other optimizations are equal.}

\label{tab:complex}
\end{table*}

\textbf{Cheap K-mer Selection (CKS)} aims to reduce seed frequencies by selecting seeds from a wider
potential seed pool. For a fixed seed length $k$, CKS samples $\lfloor \frac{L}{k} \rfloor$ seed
positions consecutively in a read, with each position apart from another by $k$-bp. Among the
$\lfloor \frac{L}{k} \rfloor$ positions, it selects $x$ seed positions that yield the least frequent
seeds (assuming the mapper needs $x$ seeds). In this way, it avoids using positions that generate
frequent seeds.

CKS has low overhead. In total, CKS only needs $\lfloor \frac{L}{k} \rfloor$ lookups for seed
frequencies followed by a sort of $\lfloor \frac{L}{k} \rfloor$ seed frequencies. Although fast, CKS
can only provide limited seed frequency reduction as it has a very limited pool to select seeds
from. For instance, in a common mapping setting where the read length $L$ is 100-bp and seed length
$k$ is 12, the read can be divided into at most $\lfloor \frac{100}{12} \rfloor = 8$ positions. With
only 8 potential positions to select from, CKS is forced to gradually select more frequent seeds
under greater seed demands. To tolerate 5 errors in this read, CKS has to select 6 seeds out of 8
potential seed positions. This implies that CKS will select the 3rd most frequent seed out of 8
potential seeds. As we have shown in Figure~\ref{fig:seed_ana}, 12-bp seeds on average have a
frequency over 172, and selecting the 3rd frequent position out of 8 potential seeds renders a high
possibility of selecting a frequent seed which has a higher frequency than average.

Similar to CKS, \textbf{Optimal Prefix Selection (OPS)} also uses fixed length seeds. However, it
allows a greater freedom of choosing seed positions. Unlike CKS, which only select seeds at
positions that are a multiple of the seed length $k$, OPS allows seeds to be selected from any
position in the read, as long as they do not overlap.

Resembling our optimal seed finding algorithm, the basis of OPS is also a dynamic programming
algorithm that implements a simpler recurrence function. The major difference between OPS and
OSS is that OPS does not need to derive the optimal length of each seed, as the seed
length is fixed to $k$ bp. This reduces the search space of optimal seeds to a single dimension,
which is only the seed placements. The worst case/average complexity of OPS is $\mathcal{O}(L\times x)$.

Compared to CKS, OPS is more complex and requires more seed frequency lookups. In return, OPS finds
less frequent seeds, especially under large seed numbers. However, with a fixed seed length, OPS
cannot further reduce the seed frequencies by extending the seeds longer.

\textbf{Spaced seeds} aim at rebalancing frequencies among patterns in the seed database by
introducing a hash function that is guided by a user-defined bit-mask. Different patterns that
are hashed into the same hash value are considered as a single ``rebalanced seed''. By carefully
designing the hashing function, which extracts base-pairs only at selected positions from a longer (e.g.,
18-bp) pattern, spaced seeds can group up frequent long patterns with infrequent long patterns and
merge them into the new ``rebalanced seeds'', which have smaller frequency variations. At runtime,
long raw seeds are selected consecutively in the reads, which are processed by the rebalancing hash
function later.

Compared to OSS, spaced seeds has two disadvantages. First, the hash function
cannot perfectly balance frequencies among all ``rebalanced seeds''. After rebalancing, there is still
a large disparity in seed frequency amongst seeds. Second, seed placement in spaced seeds is static, and does
not accommodate for high frequency seeds. Therefore, positions
that generate frequent seeds are not avoided which still give rise to the frequent seeds
phenomenon (frequent seeds are encountered more frequently in the reads, if selected consecutively).

\section{Results}

In this section, we compare the average case complexity, memory traffic and effectiveness of
OSS against the four prior studies, ASF, CKS, OPS and spaced seeds as well as the naive mechanism, which
selects fixed seeds consecutively. Memory traffic is measured by the number of required seed
frequency lookups to map a single read. The effectiveness of a seed selection scheme is measured by
the average seed frequency of mapping 4,031,354 101-bp reads from a real read set, ERR240726 from
the 1000 Genomes Project, under different numbers of seeds.

We do not measure the execution time of each mechanism because different seed selection
optimizations are combined with different seed database implementations. CKS, OPS and spaced seeds
use hash tables for short, fixed-length seeds while ASF and OSS employs slower but more memory
efficient BWT and FM-index for longer, variant-length seeds. However, this combination is
inter-changeable. CKS, OPS, and spaced seeds can also work well with BWT and FM-index and ASF and
OSS can also be combined with a large hash-table, given sufficient memory space. Besides, different
existing implementations have their unique seed database optimizations, which introduces more
variations to the execution time. With the above reasons, we only compare the complexity and memory
traffic of each seed selection scheme, without measuring their runtime performance.

We benchmark each seed optimization scheme with multiple configurations. We benchmark ASF with
multiple frequency thresholds, 5, 10, 100, 500 and 1000. If a read fails to provide enough seeds
in ASF, due to having many long seeds under small thresholds, the read will be processed again in
CKS with a fixed seed length of 12-bp. We benchmark CKS, OPS and naive under three fixed seed
lengths, 12, 13 and 14. We benchmark spaced seeds with the default bit-mask, ``110100110010101111'',
which hashes 18-bp long seeds into 11-bp long signatures.

All seed selection mechanisms are benchmarked using an in-house seed database, which supports variant
seed lengths between $S_{min}=10$ and $S_{max}=30$.

\begin{figure}[h]
  \centering
  \includegraphics[width=0.48\textwidth]{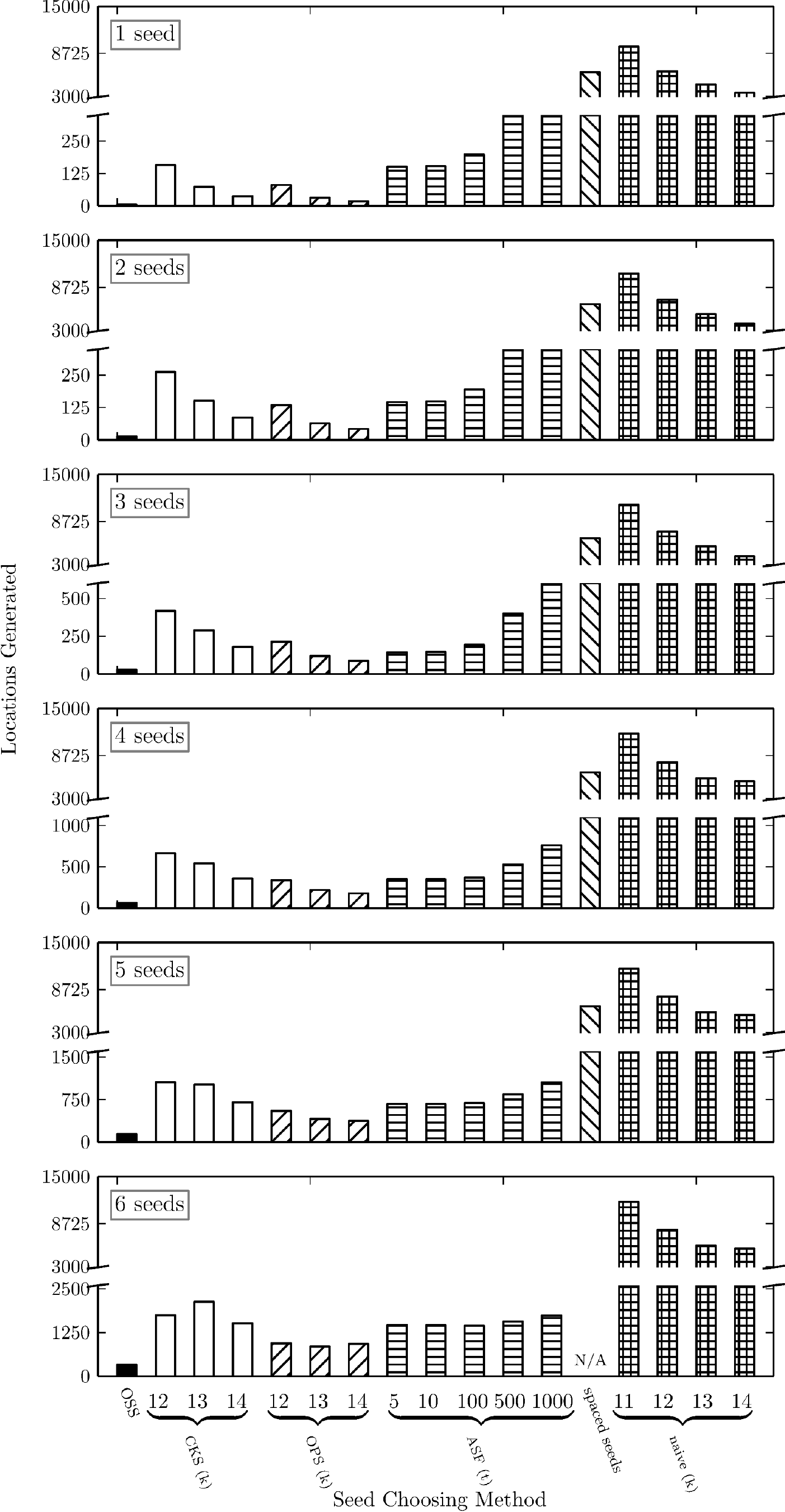}
  \caption{Average seed frequency comparison among Optimal Seed Solver (OSS), Adaptive Seeds Filter
  (ASF), Cheap K-mer Selection (CKS), Optimal Prefix Selection (OPS), spaced seeds and naive
  (selecting fixed length seeds consecutively). The results are gathered by mapping 4031354
  101-bp reads from the read set ERR240726\_1 from 1000 Genome Project under different numbers of
  seeds (for
  better accuracy and error tolerance). In each figure, a smaller average seed frequency indicates a
  more effective seed selection mechanism.}
  \label{fig:result}
\end{figure}

Table~\ref{tab:complex} summarizes the average complexity and memory traffic of each seed selection
optimization. From the table, we can observe that OSS requires the most seed frequency lookups
($\mathcal{O}(L^{2})$) with the worst average case complexity, ($\mathcal{O}(x \times L)$), which
tied with OPS. Nonetheless, OSS is the most effective seed selection scheme as
Figure~\ref{fig:result} shows. Among all seed selection optimizations, OSS provides the largest
frequency reduction of seeds on average, achieving a 3x larger frequency reduction compared to the
second best seed selection scheme, OPS.

As shown in Figure~\ref{fig:result}, the average seed frequencies of OSS, CKS and OPS increase with
more seeds. This is expected as there is less flexibility in seed placement with more seeds. For
OSS, more seeds also means shorter average seed length, which also contributes to greater average
seed frequencies. For ASF, average seed frequencies remains similar for three or fewer seeds. When
there are more than three seeds, the average seed frequencies increase with more seeds. This is
because up until three seeds, all reads have enough base-pairs to accommodate all seeds, since the
maximum seed length is $S_{max}=30$. However, once beyond three seeds, reads start to fail in ASF
(having insufficient base-pairs to accommodate all seeds) and the failed reads are passed to CKS
instead. Therefore the increase after three seeds is mainly due to the increase in CKS. For $t=10$
with six seeds, we observe from our experiment that 66.4\% of total reads fail ASF and are processed
in CKS instead.

For CKS and OPS, the average seed frequency decreases with increasing seed length when the number of
seeds is small (e.g., $<4$). When the number of seeds is large (e.g., $6$), it is not obvious if
greater seed lengths provide smaller average seed frequencies. In fact, for 6 seeds, the average
seed frequency of OPS rises slightly when we increase the seed length from 13 bp to 14 bp. This is
because, for small numbers of seeds, the read has plenty of space to arrange and accommodate the
slightly longer seeds.  Therefore, in this case, longer seeds reduce the average seed frequency.
However, for large numbers of seeds, even a small increase in seed length will significantly
decrease the flexibility in seed arrangement. In this case, the frequency reduction of longer seeds
is surpassed by the frequency increase of reduced flexibility in seed arrangement. Moreover, the
benefit of having longer seeds diminishes with greater seed lengths. Many seeds are already
infrequent at 12-bp. Extending the infrequent seeds longer does not introduce much reduction in the
total seed frequency. This result corroborates the urge of enabling flexibility in both individual
seed length and seed placements.

Overall, OSS provides the least frequent seeds on average, achieving a 3x larger frequency reduction
than the second best seed selection schemes, OPS.

\section{Discussion}

As shown in the Results section, OSS requires $\mathcal{O}(L^{2})$ seed-frequency lookups in order
to derive the optimal solution of a read. For a non-trivial seed database implementation such as BWT
with FM-index, this can be a time consuming process. For reads that do not include any frequent
short patterns, OSS can be an expensive procedure with little benefit, as simpler seed selection
mechanisms can also produce low-frequency seeds. Therefore, when designing a mapper, OSS is best
used in junction with other greedy seed selection algorithms. In such configuration, OSS will only
be invoked when greedy seed selection algorithms fail to deliver infrequent seeds. However, such
study is beyond the scope of this paper and will be explored in our future research.

The Optimal Seed Solver also revealed that there is still great potential in designing better greedy
seed selection optimizations. From our experiment, we observe that the most effective greedy seed
selection optimization still provides 3x more frequent seeds on average than optimal. Better greedy
algorithms that provides less frequent seeds without a large number of database lookups will also be
included in our future research.

\section{Conclusion}

Optimizing seed selection is an important problem in read mapping. The number of selected
non-overlapping seeds defines the error tolerance of a mapper while the total frequency of all
selected seeds determines the performance of the mapper. To build a fast while error tolerant
mapper, it is essential to select a large number of non-overlapping seeds while keeping each seed as
infrequent as possible. In this paper, we confirmed \emph{frequent seed phenomenon} discovered in
previous works~\cite{Kielbasa2011}, which suggests that in a naive seed selection scheme, mappers
tend to select frequent seeds from reads, even when using long seeds. To solve this problem, we
proposed \emph{Optimal Seed Solver} (OSS), a dynamic-programming algorithm that derives the optimal
set of seeds from a read that has the minimum total frequency. We further improved OSS with two
optimizations, \emph{optimal divider cascading} and \emph{early seed termination}. With both
optimizations, we reduced the average-case complexity of OSS to $\mathcal{O}(x \times L)$, and
achieved a $\mathcal{O}(x \times L^{2})$ worst-case complexity. We compared OSS to four prior
studies, Adaptive Seeds Filter, Cheap K-mer Selection, Optimal Prefix Selection and spaced seeds and
showed that OSS provided a 3-fold seed frequency reduction over the best previous seed selection
scheme, Optimal Prefix Selection.


\paragraph{Funding\textcolon}

This study is supported by an NIH grant (HG006004) to C. Alkan and O. Mutlu, HG007104 to C.
Kingsford) and a Marie Curie Career Integration Grant (PCIG-2011-303772) to C. Alkan under the
Seventh Framework Programme. C. Alkan also acknowledges support from The Science Academy of Turkey,
under the BAGEP program.



\begin{thebibliography}{10}

\bibitem{1000GP}
{1000 Genomes Project Consortium}.
\newblock A map of human genome variation from population-scale sequencing.
\newblock {\em Nature}, 467:1061--1073, 2010.

\bibitem{1000GP2012}
{1000 Genomes Project Consortium}.
\newblock An integrated map of genetic variation from 1,092 human genomes.
\newblock {\em Nature}, 491(7422):56--65, Nov 2012.

\bibitem{hobbes}
A.~Ahmadi, A.~Behm, N.~Honnalli, C.~Li, L.~Weng, and X.~Xie.
\newblock Hobbes: optimized gram-based methods for efficient read alignment.
\newblock {\em Nucleic Acids Research}, 40:e41, 2011.

\bibitem{Alkan2009}
C.~Alkan, J.~M. Kidd, T.~Marques-Bonet, G.~Aksay, F.~Antonacci, F.~Hormozdiari,
  J.~O. Kitzman, C.~Baker, M.~Malig, O.~Mutlu, S.~C. Sahinalp, R.~A. Gibbs, and
  E.~E. Eichler.
\newblock Personalized copy number and segmental duplication maps using
  next-generation sequencing.
\newblock {\em Nat Genet}, 41:1061--1067, 2009.

\bibitem{Burrows94ablock-sorting}
M.~Burrows, D.~J. Wheeler, M.~Burrows, and D.~J. Wheeler.
\newblock A block-sorting lossless data compression algorithm.
\newblock 1994.

\bibitem{FM-index}
P.~Ferragina and G.~Manzini.
\newblock Opportunistic data structures with applications.
\newblock In {\em Proceedings of the 41st Annual Symposium on Foundations of
  Computer Science}, FOCS '00, pages 390--, Washington, DC, USA, 2000. IEEE
  Computer Society.

\bibitem{Flannick2014}
J.~Flannick, G.~Thorleifsson, N.~L. Beer, S.~B.~R. Jacobs, N.~Grarup, N.~P.
  Burtt, A.~Mahajan, C.~Fuchsberger, G.~Atzmon, R.~Benediktsson, J.~Blangero,
  D.~W. Bowden, I.~Brandslund, J.~Brosnan, F.~Burslem, J.~Chambers, Y.~S. Cho,
  C.~Christensen, D.~A. Douglas, R.~Duggirala, Z.~Dymek, Y.~Farjoun,
  T.~Fennell, P.~Fontanillas, T.~Fors{\'{e}}n, S.~Gabriel, B.~Glaser, D.~F.
  Gudbjartsson, C.~Hanis, T.~Hansen, A.~B. Hreidarsson, K.~Hveem, E.~Ingelsson,
  B.~Isomaa, S.~Johansson, T.~J{\o}rgensen, M.~E. J{\o}rgensen, S.~Kathiresan,
  A.~Kong, J.~Kooner, J.~Kravic, M.~Laakso, J.-Y. Lee, L.~Lind, C.~M. Lindgren,
  A.~Linneberg, G.~Masson, T.~Meitinger, K.~L. Mohlke, A.~Molven, A.~P. Morris,
  S.~Potluri, R.~Rauramaa, R.~Ribel-Madsen, A.-M. Richard, T.~Rolph,
  V.~Salomaa, A.~V. Segr{\`{e}}, H.~Sk{\"{a}}rstrand, V.~Steinthorsdottir,
  H.~M. Stringham, P.~Sulem, E.~S. Tai, Y.~Y. Teo, T.~Teslovich,
  U.~Thorsteinsdottir, J.~K. Trimmer, T.~Tuomi, J.~Tuomilehto, F.~Vaziri-Sani,
  B.~F. Voight, J.~G. Wilson, M.~Boehnke, M.~I. McCarthy, P.~R. Nj{\o}lstad,
  O.~Pedersen, G.-T.~C. ~, T.-G. E. N. E. S.~C. ~, L.~Groop, D.~R. Cox,
  K.~Stefansson, and D.~Altshuler.
\newblock Loss-of-function mutations in SLC30A8 protect against type 2
  diabetes.
\newblock {\em Nature Genetics}, 46(4):357--363, Apr 2014.

\bibitem{Flicek2009-mapping}
P.~Flicek and E.~Birney.
\newblock {Sense from sequence reads: methods for alignment and assembly.}
\newblock {\em Nature methods}, 6(11 Suppl):S6--S12, Nov. 2009.

\bibitem{Green2010}
R.~E. Green, J.~Krause, A.~W. Briggs, T.~Maricic, U.~Stenzel, M.~Kircher,
  N.~Patterson, H.~Li, W.~Zhai, M.~H.-Y. Fritz, N.~F. Hansen, E.~Y. Durand,
  A.-S. Malaspinas, J.~D. Jensen, T.~Marques-Bonet, C.~Alkan, K.~Pr�fer,
  M.~Meyer, H.~A. Burbano, J.~M. Good, R.~Schultz, A.~Aximu-Petri, A.~Butthof,
  B.~H�ber, B.~H�ffner, M.~Siegemund, A.~Weihmann, C.~Nusbaum, E.~S.
  Lander, C.~Russ, et~al.
\newblock A draft sequence of the {N}eandertal genome.
\newblock {\em Science}, 328:710--722, 2010.

\bibitem{Kielbasa2011}
S.~Kie\l{}basa, R.~Wan, K.~Sato, P.~Horton, and M.~C. Frith.
\newblock Adaptive seeds tame genomic sequence comparison.
\newblock {\em Genome Research}, 21(3):487--493, 2011.

\bibitem{patternhunter}
B.~Ma, J.~Tromp, and M.~Li.
\newblock Patternhunter: faster and more sensitive homology search.
\newblock {\em Bioinformatics}, 18:440--445, 2002.

\bibitem{Marco-Sola2012}
S.~Marco-Sola, M.~Sammeth, R.~Guig�, and P.~Ribeca.
\newblock The gem mapper: fast, accurate and versatile alignment by filtration.
\newblock {\em Nat Methods}, 9(12):1185--1188, 2012.

\bibitem{Marques-Bonet2009}
T.~Marques-Bonet, J.~M. Kidd, M.~Ventura, T.~A. Graves, Z.~Cheng, L.~W.
  Hillier, Z.~Jiang, C.~Baker, R.~Malfavon-Borja, L.~A. Fulton, C.~Alkan,
  G.~Aksay, S.~Girirajan, P.~Siswara, L.~Chen, M.~F. Cardone, A.~Navarro, E.~R.
  Mardis, R.~K. Wilson, and E.~E. Eichler.
\newblock A burst of segmental duplications in the genome of the {African}
  great ape ancestor.
\newblock {\em Nature}, 457:877--881, 2009.

\bibitem{Meyer2012}
M.~Meyer, M.~Kircher, M.-T. Gansauge, H.~Li, F.~Racimo, S.~Mallick, J.~G.
  Schraiber, F.~Jay, K.~Pr{\"{u}}fer, C.~{de Filippo}, P.~H. Sudmant, C.~Alkan,
  Q.~Fu, R.~Do, N.~Rohland, A.~Tandon, M.~Siebauer, R.~E. Green, K.~Bryc, A.~W.
  Briggs, U.~Stenzel, J.~Dabney, J.~Shendure, J.~Kitzman, M.~F. Hammer, M.~V.
  Shunkov, A.~P. Derevianko, N.~Patterson, A.~M. Andr{\'{e}}s, E.~E. Eichler,
  M.~Slatkin, D.~Reich, J.~Kelso, and S.~P{\"{a}}{\"{a}}bo.
\newblock A high-coverage genome sequence from an archaic Denisovan individual.
\newblock {\em Science}, 338(6104):222--226, Oct 2012.

\bibitem{Myers1999}
G.~Myers.
\newblock A fast bit-vector algorithm for approximate string matching based on
  dynamic programming.
\newblock {\em J. ACM}, 46(3):395--415, 1999.

\bibitem{nw}
S.~B. Needleman and C.~D. Wunsch.
\newblock A general method applicable to the search for similarities in the
  amino acid sequence of two proteins.
\newblock {\em Journal of Molecular Biology}, 48:443--453, 1970.

\bibitem{Ng2010}
S.~B. Ng, A.~W. Bigham, K.~J. Buckingham, M.~C. Hannibal, M.~J. McMillin, H.~I.
  Gildersleeve, A.~E. Beck, H.~K. Tabor, G.~M. Cooper, H.~C. Mefford, C.~Lee,
  E.~H. Turner, J.~D. Smith, M.~J. Rieder, K.-I. Yoshiura, N.~Matsumoto,
  T.~Ohta, N.~Niikawa, D.~A. Nickerson, M.~J. Bamshad, and J.~Shendure.
\newblock Exome sequencing identifies {MLL2} mutations as a cause of Kabuki
  syndrome.
\newblock {\em Nat Genet}, 42(9):790--793, Sep 2010.

\bibitem{Prado-Martinez2013}
J.~Prado-Martinez, P.~H. Sudmant, J.~M. Kidd, H.~Li, J.~L. Kelley,
  B.~Lorente-Galdos, K.~R. Veeramah, A.~E. Woerner, T.~D. O'Connor,
  G.~Santpere, A.~Cagan, C.~Theunert, F.~Casals, H.~Laayouni, K.~Munch,
  A.~Hobolth, A.~E. Halager, M.~Malig, J.~Hernandez-Rodriguez,
  I.~Hernando-Herraez, K.~Pr{\"{u}}fer, M.~Pybus, L.~Johnstone, M.~Lachmann,
  C.~Alkan, D.~Twigg, N.~Petit, C.~Baker, F.~Hormozdiari, M.~Fernandez-Callejo,
  M.~Dabad, M.~L. Wilson, L.~Stevison, C.~Camprub{\'{\i}}, T.~Carvalho,
  A.~Ruiz-Herrera, L.~Vives, M.~Mele, T.~Abello, I.~Kondova, R.~E. Bontrop,
  A.~Pusey, F.~Lankester, J.~A. Kiyang, R.~A. Bergl, E.~Lonsdorf, S.~Myers,
  M.~Ventura, P.~Gagneux, D.~Comas, H.~Siegismund, J.~Blanc, L.~Agueda-Calpena,
  M.~Gut, L.~Fulton, S.~A. Tishkoff, J.~C. Mullikin, R.~K. Wilson, I.~G. Gut,
  M.~K. Gonder, O.~A. Ryder, B.~H. Hahn, A.~Navarro, J.~M. Akey,
  J.~Bertranpetit, D.~Reich, T.~Mailund, M.~H. Schierup, C.~Hvilsom, A.~M.
  Andr{\'{e}}s, J.~D. Wall, C.~D. Bustamante, M.~F. Hammer, E.~E. Eichler, and
  T.~Marques-Bonet.
\newblock Great ape genetic diversity and population history.
\newblock {\em Nature}, 499(7459):471--475, Jul 2013.

\bibitem{Reich2010}
D.~Reich, R.~E. Green, M.~Kircher, J.~Krause, N.~Patterson, E.~Y. Durand,
  B.~Viola, A.~W. Briggs, U.~Stenzel, P.~L.~F. Johnson, T.~Maricic, J.~M. Good,
  T.~Marques-Bonet, C.~Alkan, Q.~Fu, S.~Mallick, H.~Li, M.~Meyer, E.~E.
  Eichler, M.~Stoneking, M.~Richards, S.~Talamo, M.~V. Shunkov, A.~P.
  Derevianko, J.-J. Hublin, J.~Kelso, M.~Slatkin, and S.~P��bo.
\newblock Genetic history of an archaic hominin group from {Denisova Cave} in
  {Siberia}.
\newblock {\em Nature}, 468:1053--1060, 2010.

\bibitem{Rognes11}
T.~Rognes.
\newblock Faster smith-waterman database searches with inter-sequence simd
  parallelisation.
\newblock {\em BMC Bioinformatics}, 12:221, 2011.

\bibitem{shrimp}
S.~M. Rumble, P.~Lacroute, A.~V. Dalca, M.~Fiume, A.~Sidow, and M.~Brudno.
\newblock SHRiMP: Accurate mapping of short color-space reads.
\newblock {\em PLoS Comput Biol}, 5:e1000386, 2009.

\bibitem{Scally2012}
A.~Scally, J.~Y. Dutheil, L.~W. Hillier, G.~E. Jordan, I.~Goodhead, J.~Herrero,
  A.~Hobolth, T.~Lappalainen, T.~Mailund, T.~Marques-Bonet, S.~McCarthy, S.~H.
  Montgomery, P.~C. Schwalie, Y.~A. Tang, M.~C. Ward, Y.~Xue, B.~Yngvadottir,
  C.~Alkan, L.~N. Andersen, Q.~Ayub, E.~V. Ball, K.~Beal, B.~J. Bradley,
  Y.~Chen, C.~M. Clee, S.~Fitzgerald, T.~A. Graves, Y.~Gu, P.~Heath, A.~Heger,
  et~al.
\newblock Insights into hominid evolution from the gorilla genome sequence.
\newblock {\em Nature}, 483:169--175, 2012.

\bibitem{sw}
T.~F. Smith and M.~S. Waterman.
\newblock Identification of common molecular subsequences.
\newblock {\em Journal of Molecular Biology}, 147:195--195, 1981.

\bibitem{swps3}
A.~Szalkowski, C.~Ledergerber, P.~Krahenbuhl, and C.~Dessimoz.
\newblock {SWPS3} - fast multi-threaded vectorized {Smith-Waterman} for {IBM}
  {Cell/B}.e. and {x86/SSE2}.
\newblock {\em BMC Research Notes}, 1(1):107+, 2008.

\bibitem{Ventura2011}
M.~Ventura, C.~R. Catacchio, C.~Alkan, T.~Marques-Bonet, S.~Sajjadian, T.~A.
  Graves, F.~Hormozdiari, A.~Navarro, M.~Malig, C.~Baker, C.~Lee, E.~H. Turner,
  L.~Chen, J.~M. Kidd, N.~Archidiacono, J.~Shendure, R.~K. Wilson, and E.~E.
  Eichler.
\newblock Gorilla genome structural variation reveals evolutionary parallelisms
  with chimpanzee.
\newblock {\em Genome Res}, 21:1640--1649, 2011.

\bibitem{razers3}
D.~Weese, M.~Holtgrewe, and K.~Reinert.
\newblock {RazerS} 3: Faster, fully sensitive read mapping.
\newblock {\em Bioinformatics}, 28(20):2592--2599, 2012.

\bibitem{SHD}
H.~Xin, J.~Greth, J.~Emmons, G.~Pekhimenko, C.~Kingsford, C.~Alkan, and
  O.~Mutlu.
\newblock Shifted hamming distance: A fast and accurate simd-friendly filter to
  accelerate alignment verification in read mapping.
\newblock {\em Bioinformatics}, Jan 2015.

\bibitem{Xin2013}
H.~Xin, D.~Lee, F.~Hormozdiari, S.~Yedkar, O.~Mutlu, and C.~Alkan.
\newblock Accelerating read mapping with {FastHASH}.
\newblock {\em BMC Genomics}, 14(Suppl 1):S13, 2013.


\end{thebibliography}

\clearpage
\newpage
\setcounter{section}{0}

\begin{figure}[h] \centering
\includegraphics[width=0.48\textwidth]{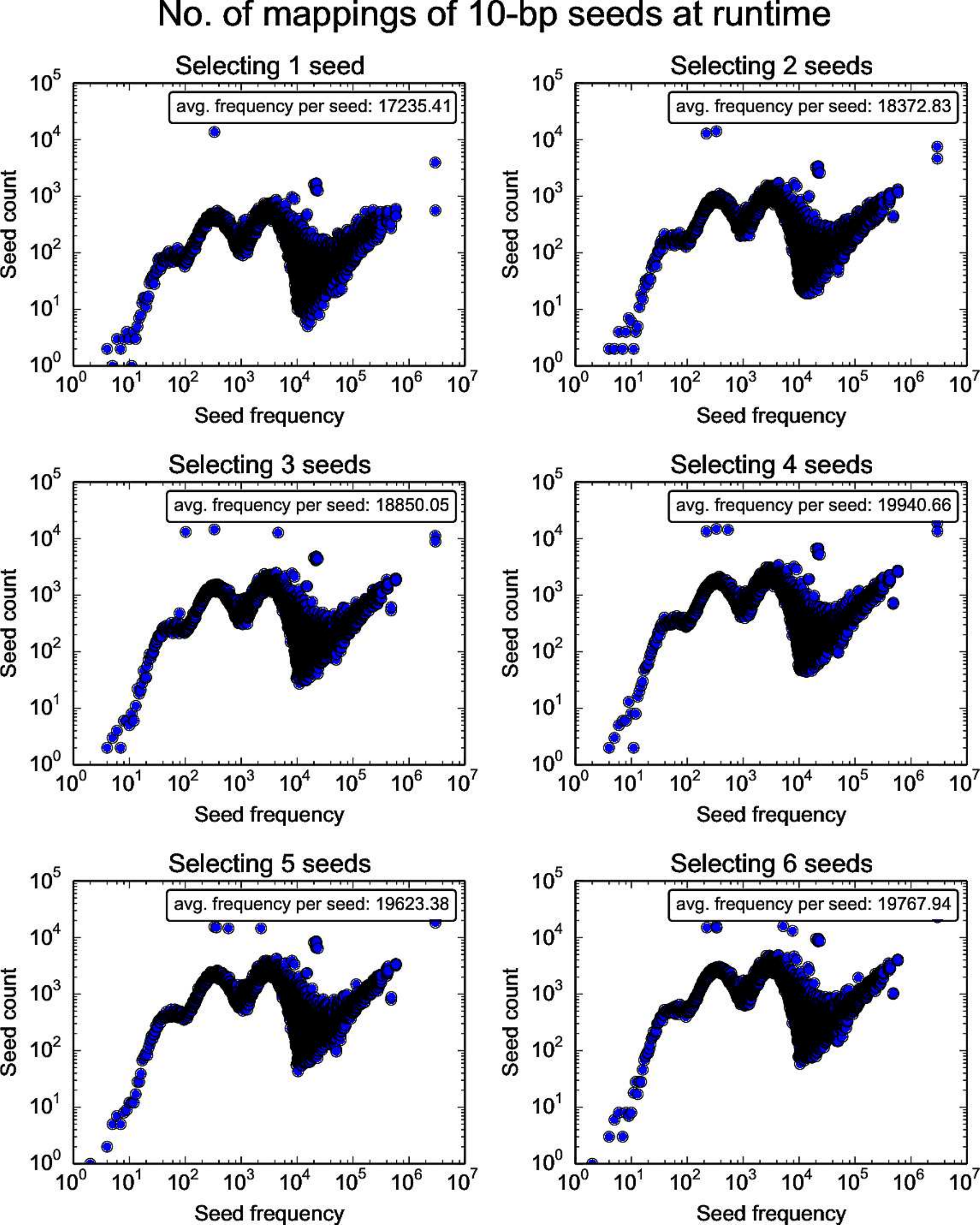}
\caption{Frequency distribution of 10-bp seeds at runtime by selecting seed consecutively under different
number of required seeds.}
\label{fig:seed_10}
\end{figure}

\begin{figure}[h] \centering
\includegraphics[width=0.48\textwidth]{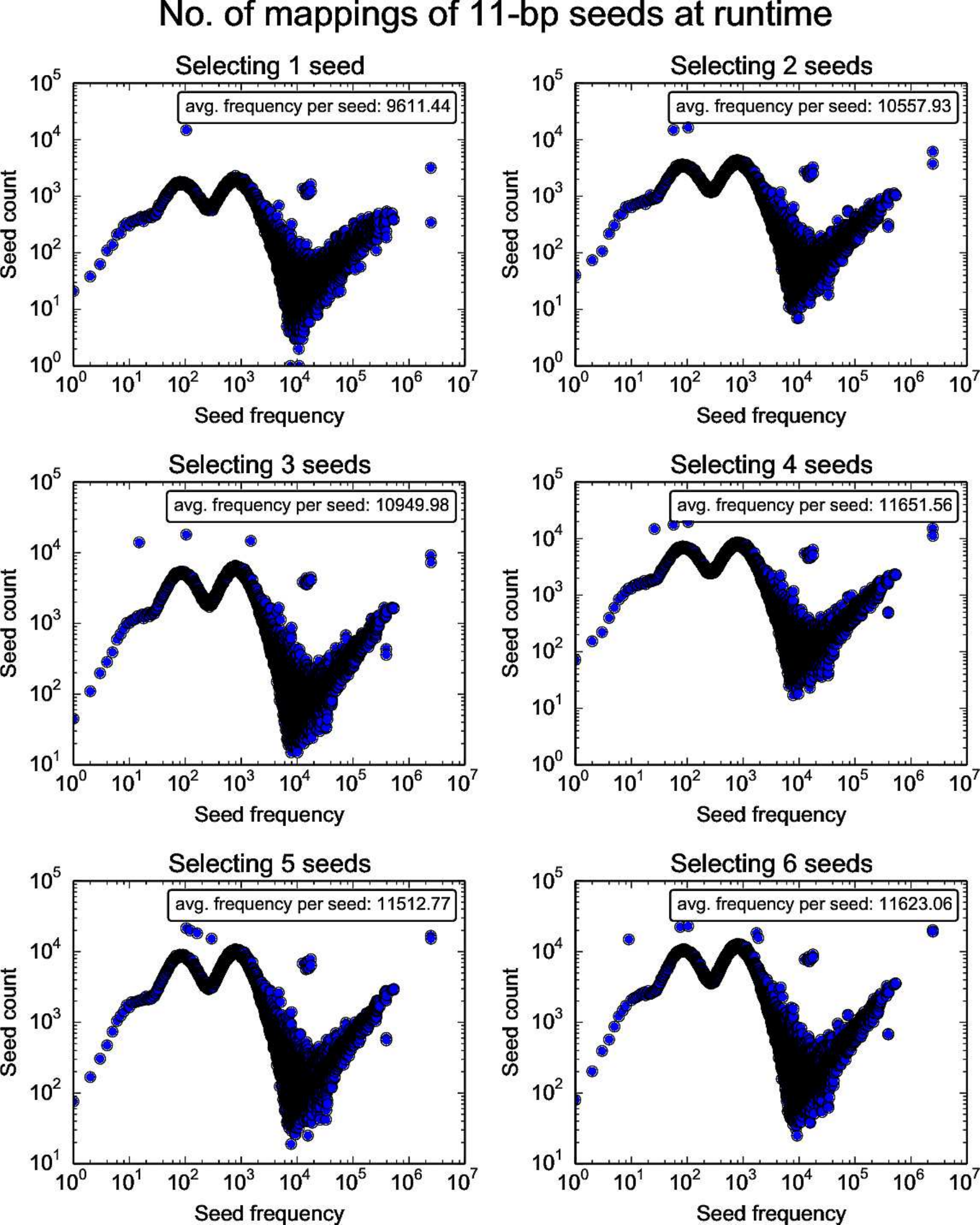}
\caption{Frequency distribution of 11-bp seeds at runtime by selecting seed consecutively under different
number of required seeds.}
\label{}
\end{figure}

\begin{figure}[h] \centering
\includegraphics[width=0.48\textwidth]{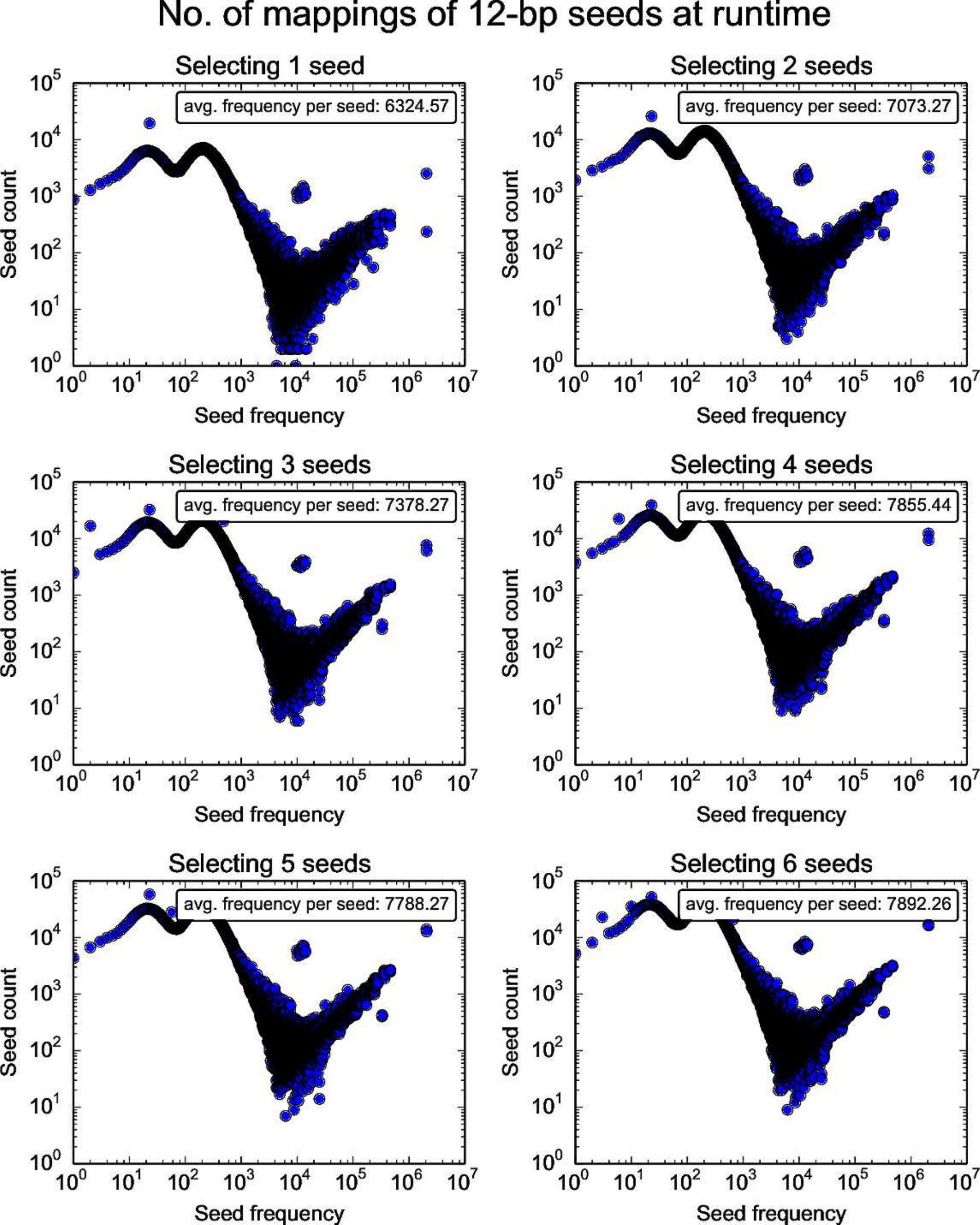}
\caption{Frequency distribution of 12-bp seeds at runtime by selecting seed consecutively under different
number of required seeds.}
\label{}
\end{figure}

\begin{figure}[h] \centering
\includegraphics[width=0.48\textwidth]{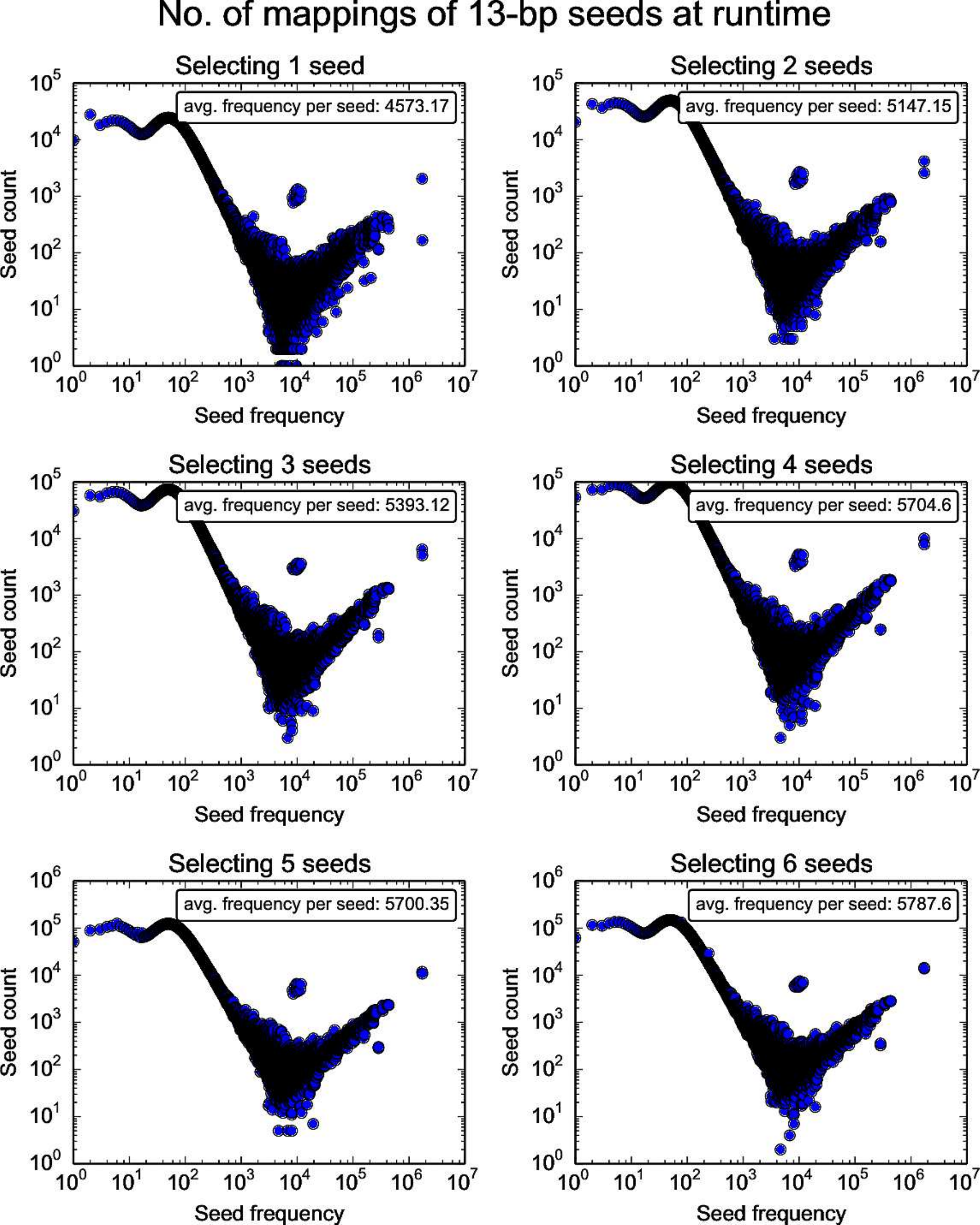}
\caption{Frequency distribution of 13-bp seeds at runtime by selecting seed consecutively under different
number of required seeds.}
\label{}
\end{figure}

\begin{figure}[h] \centering
\includegraphics[width=0.48\textwidth]{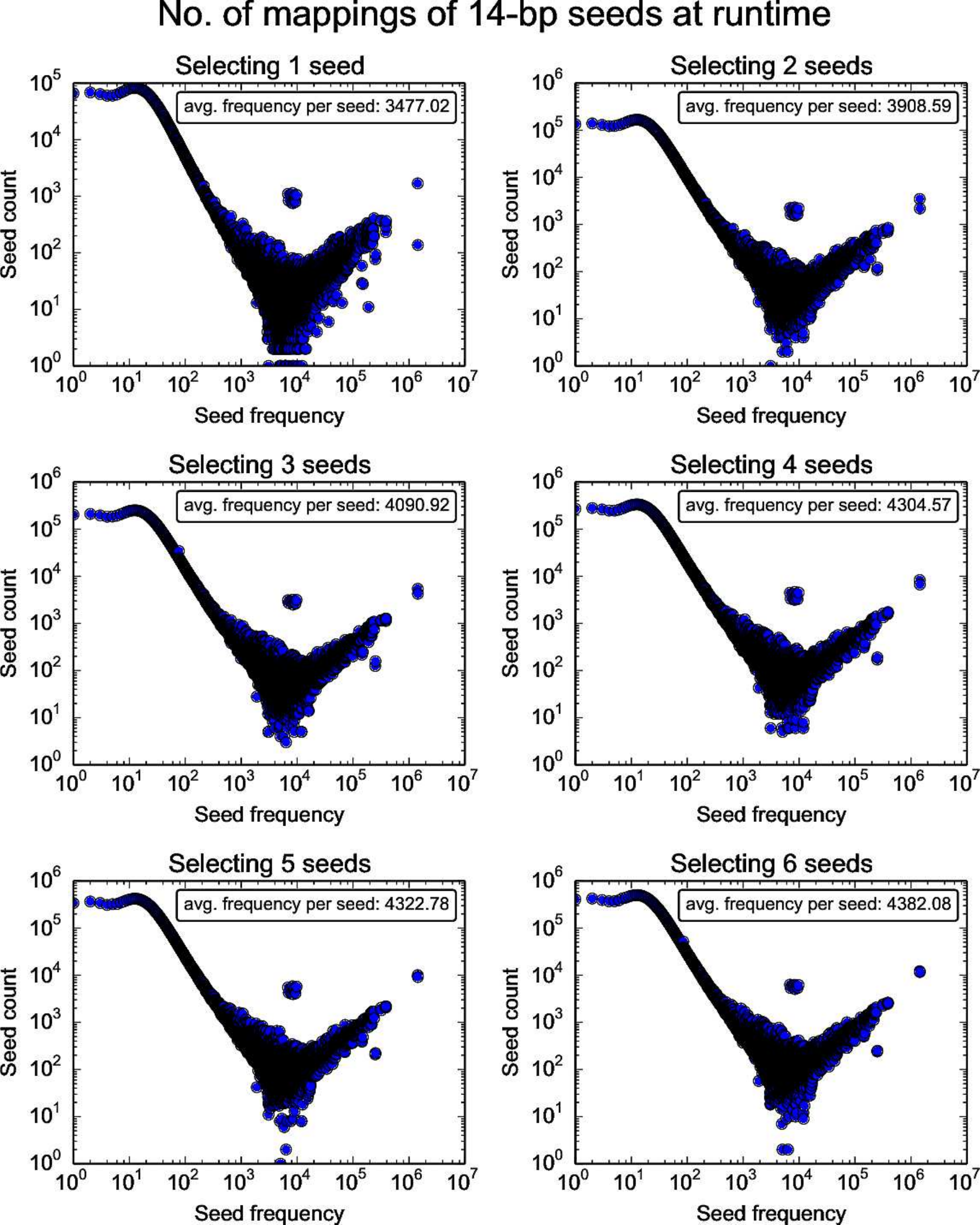}
\caption{Frequency distribution of 14-bp seeds at runtime by selecting seed consecutively under different
number of required seeds.}
\label{fig:seed_14}
\end{figure}

\clearpage

\section {Supplementary Materials}

\subsection{Runtime frequency distributions of seeds under variant lengths}

This section presents seed frequency distributions at runtime with regard to different seed lengths.
The results are obtained by naively selecting different fixed-length seeds consecutively in the
process of mapping 4,031,354 101-bp reads from a real read set, ERR240726, from the 1000 Genome
Project.

Figure~\ref{fig:seed_10} to Figure~\ref{fig:seed_14} from page 11-13 show seed frequency distributions of
fixed-length seeds from 10-bp to 14-bp. From these figures, we have three observations: (1) the
average seed frequencies of longer seeds are smaller, (2) beyond the seed frequency of $10^4$, more
frequent seeds are more frequently selected at runtime and (3) compared to
Figure~\ref{fig:seed_ana}, the average frequencies of selected seeds are much larger than the average
frequencies of seeds with equal length in the seed database.

As shown in all five figures above, after $10^{4}$, the seed count increases with greater seed
frequencies, which implies that frequent seeds are often selected from reads, regardless of the seed
length.

\subsection{Proof of optimal divider cascading}

This section presents the detailed proof of the optimal divider cascading phenomenon.

The optimal divider cascading phenomenon can be explained with two \emph{lemmas}:
\begin{lemma}
For any two substrings from the same iteration in OSS, one substring must include the other. Among
the two substrings, the minimum seed frequency of the outer substring must not be greater than the
minimum seed frequency of the inner substring.
\label{lem:inclusion}
\end{lemma}

The proof of Lemma~\ref{lem:inclusion} is provided below:

\begin{proof}
Since all substrings in the ``Optimal Seed Solver'' algorithm start at the beginning of the read,
any two substrings from the same iteration must have one include another, as shown in
Figure~\ref{fig:cascade}.

We prove the second part of the lemma using contradiction. Assume the outer substring has a greater
\emph{optimal frequency} (total seed frequency of the optimal seeds) than the inner substring.
Because the inner substring is included by the outer substring, the optimal seeds of the inner
substring are also valid seeds for the outer substring. Yet, the total frequency of this particular
set of seeds is smaller than the optimal frequency of the outer substring, which leads to a
contradiction.
\qed
\end{proof}

\begin{figure}[h]
  \centering
  \includegraphics[width=0.48\textwidth]{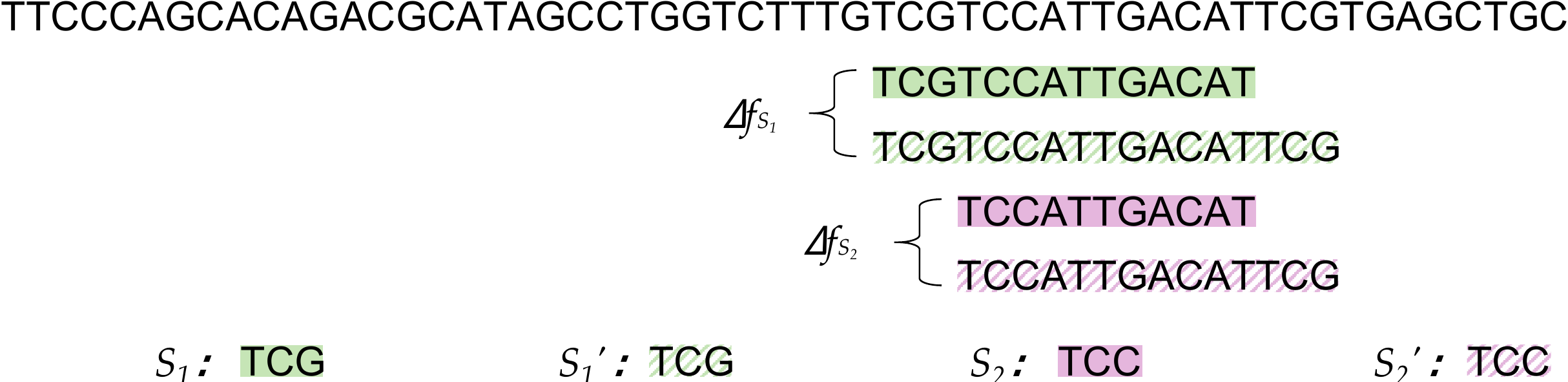}
  \caption{This figure shows two seeds $S_1$ and $S_2$, which are taken from the same read and end at
  the same position, with $S_1$ including $S_2$. Both seeds are extended by 3-bp into $S_1'$ and $S_2'$
  respectively.}
  \label{fig:delta}
\end{figure}

\begin{lemma}
When extending two \textbf{seeds} of different lengths that \textbf{end} at the same position in the
read by equal numbers of base-pairs, as one seed includes the other as shown in
Figure~\ref{fig:delta}, the frequency reduction ($\Delta f$) of extending the outer seed ($S_2
\rightarrow S_2'$) must not be greater than the frequency reduction of extending the inner seed
($S_1 \rightarrow S_1'$).
\label{lem:seeds}
\end{lemma}

\begin{figure}[h]
  \centering
  \includegraphics[width=0.30\textwidth]{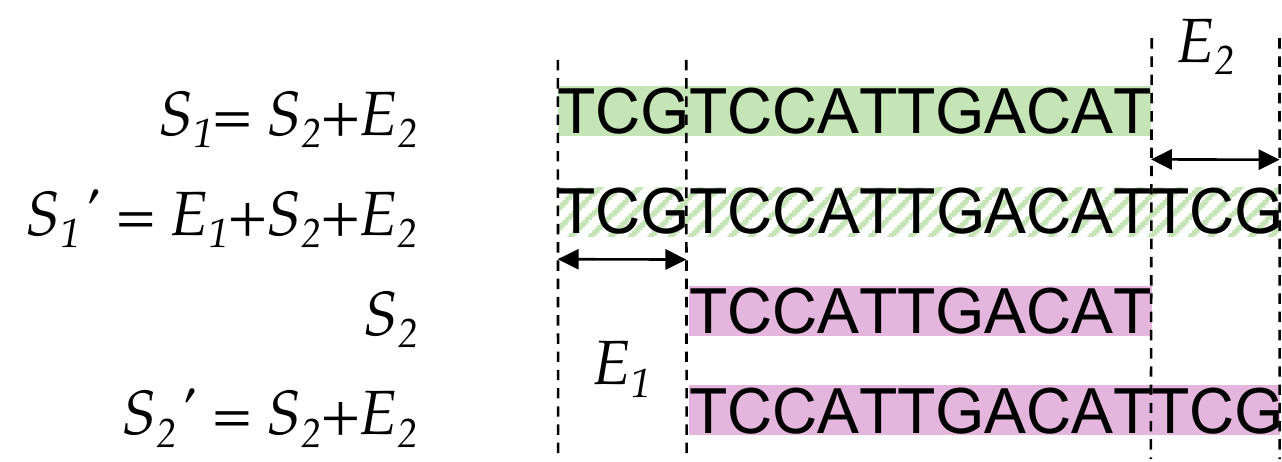}
  \caption{Two seeds, $S_1$ and $S_2$ are taken from the same read and end at the same position in
  the read. Both $S_1$ and $S_2$ are extended by 3 bp into $S_1'$ and $S_2'$ respectively.
  Considering $S_1$ as a left-extension of $S_2$ by $E_1$ and $S_1'$ as a right-extension of $S_1$
  by $E_2$, then we have $S_1 = E_1 + S_2 + E_2$.}
  \label{fig:extend}
\end{figure}

Lemma~\ref{lem:seeds} can be proven with the monotonic non-increasing property of seed frequency with
regard to a greater seed length. For example, in Figure~\ref{fig:extend}, there are two seeds taken
from the same read, $S_1$ and $S_2$, with $S_1$ including $S_2$ and both end at the same position in the
read. Now, we simultaneously extend both $S_1$ and $S_2$ longer in the read (by taking more bp) by
$3$ bp, into $S_1'$ and $S_2'$ respectively. With $\Delta f$ denoting the change of seed frequencies
before and after extension, we can claim that $\Delta f_{S_1} \leq \Delta f_{S_2}$.

To prove this inequality, it is essential to understand how is $\Delta f$ calculated. As
Figure~\ref{fig:extend} also shows, among the two seeds $S_1$ and $S_2$, $S_1$ can be considered as
a ``left-extension'' of $S_2$. Therefore, $S_1$ can be represented as $S_1=E_1+S_2$, where $E_1$
denotes the left extension of $S_1$ and the ``$+$'' sign denotes a concatenation of strings.
Similarly, $S_1'$ can be represented as a ``right-extension'' of $S_1$, which can be also written as
$S_1'=E_1+S_2+E_2$, where $E_2$ is the right $m$-bp extension of $S_1$. By the same token, we also
have $S_2'=S_2+E_2$. If $\freq{S}$ denotes the frequency of a seed $S$, then $\Delta
f_{S_1}=\freq{S_1}-\freq{S_1'}=\freq{E_1+S_2}-\freq{E_1+S_2+E_2}$.

Below, we provide the proof of Lemma~\ref{lem:seeds}:

\begin{proof} If \textbf{set} $\overline{\mathbb{E}_2}$ denotes \textbf{all} DNA sequences that are
equal in length with $E_2$ but excludes $E_2$ itself, which can be written as
$\overline{\mathbb{E}_2}=\{s\;|\;(s\in \mathrm{DNA\ sequence})\wedge (|s| = |E_2|)\wedge (s\neq
E_2)\}$,
then the reduced frequency of $S_1$ and $S_2$ can also be written as: $$\Delta f_{S_1}=\sum_{s \in
\overline{\mathbb{E}_2}} \freq{E_1+S_2+s}$$ $$\Delta f_{S_2}=\sum_{s \in \overline{\mathbb{E}_2}}
\freq{S_2+s}$$ The right hand side of both equations denote the sum of frequencies of all seeds that
share the same beginning sequence $E_1+S_2$ (or just $S_2$) other than the sequence $E_1+S_2+E_2$
itself (or $S_2+E_2$ for $S_2'$), which is indeed $\freq{E_1+S_2}-\freq{E_1+S_2+E_2}$ (or
$\freq{S_2}-\freq{S_2+E_2}$ for $S_2$).

From both equations, we can see that both $\Delta f_{S_1}$ and $\Delta f_{S_2}$ iterates through the
same set of strings, $\overline{\mathbb{E}_2}$. For each string $i$ in set
$\overline{\mathbb{E}_2}$, we have $\freq{E_1+S_2+i} \leq \freq{S_2+i}$, as the extended longer seed
can only be less or equally frequent as the original shorter seed. Therefore, we have $\Delta
f_{S_1} \leq \Delta f_{S_2}$.
\qed
\end{proof}

From Lemma~\ref{lem:seeds}, we can deduce Corollary~\ref{cor:seeds}:
\begin{corollary}
When extending two substrings of different lengths that ends at the same position in the read by
equal number of seeds, as one substring includes the other, the frequency reduction of the
\textbf{optimal seed} (the optimal single seed) of extending the longer substring, is strictly not
greater than the frequency reduction of the optimal seed of extending the shorter substring.
\label{cor:seeds}
\end{corollary}

\begin{figure}[h]
  \centering
  \includegraphics[width=0.30\textwidth]{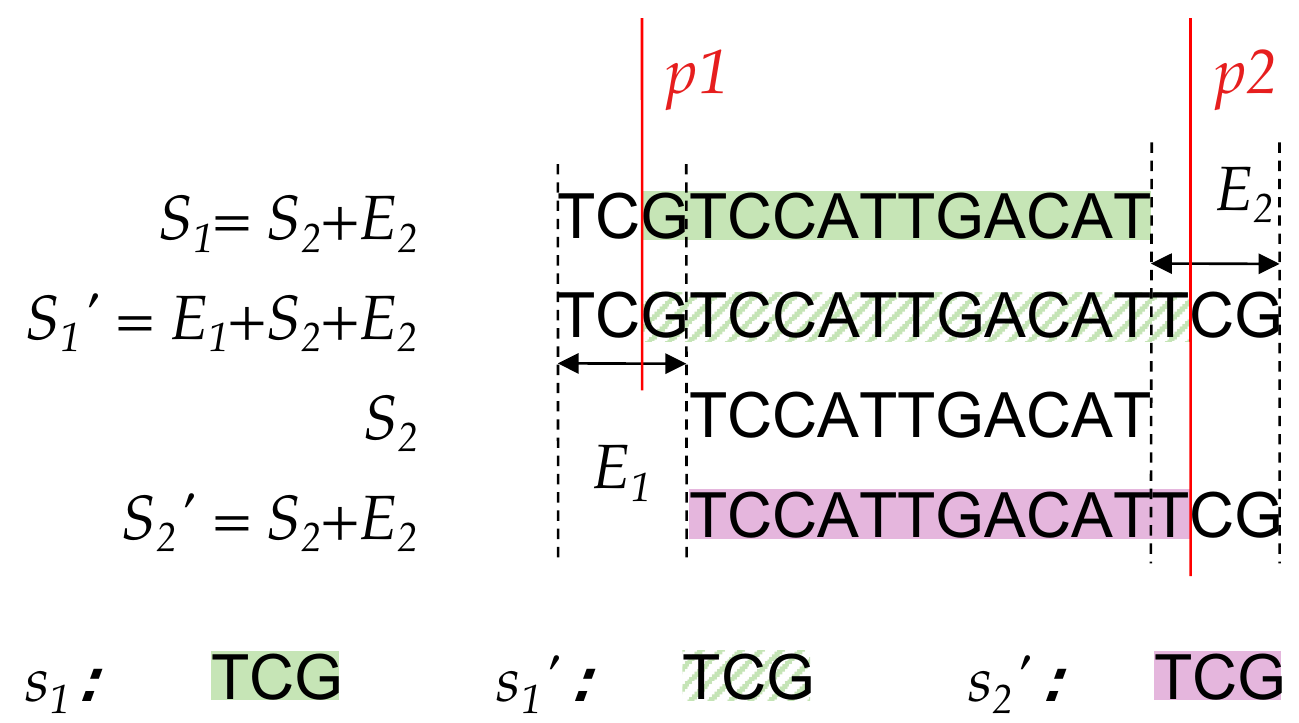}
  \caption{Among the four substrings, $S_1$, $S_2$, $S_1'$ and $S_2'$ from Figure~\ref{fig:extend},
  assume $s_1'$ is the optimal seed of substring $S_1'$. Also assume two new seeds, $s_1$ and $s_2'$.
  Between the two seeds, $s_1$ starts at where $s_1'$ starts but ends at where $S_1$ ends while $s_2'$ starts at
  where $S_2'$ starts and ends at where $s_1'$ ends.}
\label{fig:corollary}
\end{figure}

We prove Corollary~\ref{cor:seeds} by cases:

\begin{proof}
Considering the four substrings from Figure~\ref{fig:extend}, $S_1$, $S_2$, $S_1'$ and $S_2'$. Among
the four substrings, we have the following relationships:

\[ \left\{
\begin{array}{l}
	S_1 = E_1 + S_2 \\
	S_1' = E_1 + S_2 + E_2 \\
	S_2' = S_2 + E_2 \\
\end{array} \right.\]
There are in total three possible cases of where the optimal seed is selected in $S_1'$: (1) it is
selected from the region of $S_2 + E_2$, (2) it is selected from the region $E_1 + S_2$ and the optimal
seed overlaps with $E_1$ and (3) it is selected from the region of $E_1 + S_2 + E_2$ and the seed
overlaps with both $E_1$ and $E_2$. Below we prove that the Corollary is correct in each case.

Case 1: The optimal seeds is selected exclusively from $S_2 + E_2$.

This suggests that the optimal seed in $S_1'$ is also the optimal seed in $S_2'$. Based on
Lemma~\ref{lem:inclusion}, we know the optimal frequency of $S_1$ is not greater than $S_2$.

Combining the two deductions above, we can conclude that extending $S_2$ to $S_2'$ provides a
strictly no smaller frequency reduction of the optimal seed than extending $S_1$ to $S_1'$.

Case 2: The optimal seed is selected from the region $E_1 + S_2$ and it overlaps with $E_1$.

Since the optimal seed does not overlap with $E_2$, the optimal seed in both $S_1$ and $S_1'$
must be the same. Therefore extending $S_1$ to $S_1'$ provides 0 frequency reduction of the optimal
seed. As Lemma~\ref{lem:inclusion} suggests, the optimal seed frequency of $S_2$ must not be greater
than the optimal seed frequency of $S_2'$. As a result, the Corollary holds in this case.

Case 3: The optimal seed is selected across $E_1 + S_2 + E_2$ and it overlaps with both $E_1$ and $E_2$.

Assume the optimal seed, $s_1'$, in $S_1'$ starts at position $p_1$ and ends at position $p_2$.
Assume a seed, $s_1$, which starts at $p_1$ but ends where $S_1$ ends, as shown in
Figure~\ref{fig:corollary}. Also assume a seed, $s2'$, which starts at where $S_2'$ starts and ends
at $p_2$. From Lemma~\ref{lem:seeds}, we know that the reduction of seed frequency of extending
$s_1$ to $s_1'$ is no greater than the seed frequency reduction of extending $S_2$ to $s_2'$. We
also know that the optimal seed frequency of $S_1$ is no greater than the seed frequency of $s_1$
and the optimal seed frequency of $S_2'$ is no greater than the seed frequency of $s_2'$. As a
result, the frequency reduction of the optimal seed by extending $S_1$ to $S_1'$, is strictly no
greater than the frequency reduction of the optimal seed by extending $S_2$ to $S_2'$.
\qed
\end{proof}

Using Lemma~\ref{lem:inclusion}, Lemma~\ref{lem:seeds} and Corollary~\ref{cor:seeds}, we are ready
to prove that the optimal divider cascading phenomenon is always true.

\begin{theorem}
For two substrings from the same iteration in OSS, as one substring includes the other, the
\emph{first} optimal divider of the outer substring must not be at the same or a prior position than
the first optimal divider of the inner substring.
\label{theorem:cascade}
\end{theorem}

Theorem~\ref{theorem:cascade} can be proven by contradiction. The proof is provided below:

\begin{proof}
Assume $T_1$ and $T_2$ are two substrings from the same iteration in ``Optimal Seed Solver'', with $T_1$
including $T_2$. Also assume $T_1$'s first optimal divider, $D_1$, is closer to the beginning of the
read than $T_2$'s first optimal divider, $D_2$, as shown in Figure~\ref{fig:cas_proof} ($D_1 <
D_2$).

\begin{figure}[h]
  \centering
  \includegraphics[width=0.48\textwidth]{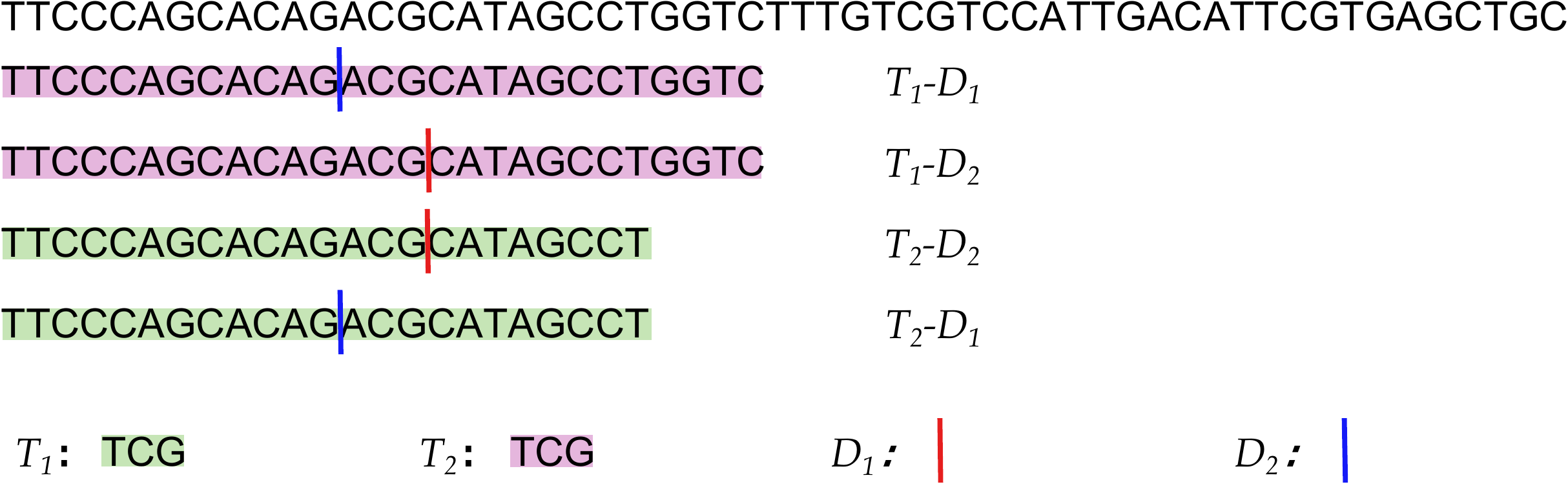}
  \caption{Two substring $T_1$ and $T_2$ taken from the same iteration. We assume $T_1$'s first optimal
  divider is at $D_1$ and $T_2$'s first optimal divider is at $D_2$.}
  \label{fig:cas_proof}
\end{figure}

Suppose we apply both divisions $D_1$ and $D_2$ to both substrings $T_1$ and $T_2$, which renders
four divisions: $T_1$-$D_1$, $T_1$-$D_2$, $T_2$-$D_1$ and $T_2$-$D_2$, as Figure~\ref{fig:cas_proof}
shows. We can prove that $T_2$-$D_2$ is a \textbf{strictly less frequent} solution than $T_2$-$D_1$.
Since $D_2$ is the first optimal divider of $T_2$ and $D_1 < D_2$, the minimum frequency of dividing
$T_2$ at $D_1$ must be greater than dividing $T_2$ at $D_2$. Let $\freq{T, D}$ denotes the optimal
frequency of dividing substring $T$ at position $D$, then based on our assumptions and lemma 2, we
have the following relationships:

\[ \left\{
\begin{array}{l}
	\freq{T_1, D_1} \leq \freq{T_1, D_2} \\
	\freq{T_2, D_2} < \freq{T_2, D_1} \\
	\freq{T_1, D_1} \geq \freq{T_2, D_1} \\
	\freq{T_1, D_2} \geq \freq{T_2, D_2}
\end{array} \right.\]

Based on corollary~\ref{cor:seeds}, we know that the frequency reduction of extending $T_2$-$D_1$ to
$T_1$-$D_1$ is strictly not greater than the frequency reduction of extending $T_2$-$D_2$ to
$T_1$-$D_2$.  From Figure~\ref{fig:cas_proof}, we can observe that only the second parts of both
$T_2$-$D_1$ and $T_2$-$D_2$ are extended into $T_1$-$D_1$ and $T_1$-$D_2$ respectively. Between
$T_2$-$D_1$ and $T_2$-$D_2$, we can see that $D_1$ produces a longer second part than $D_2$. Based
on the corollary of lemma 3, the frequency reduction of extending $T_2$-$D_2$ to $T_1$-$D_2$ is no
less than the frequency reduction of extending $T_2$-$D_1$ to $T_1$-$D_1$.  Given that $\freq{T_2,
D_2} < \freq{T_2, D_1}$ from above, we prove that $\freq{T_1, D_2} < \freq{T_1, D_1}$, which
contradicts our assumption that $\freq{T_1, D_2} \geq \freq{T_2, D_2}$.  Therefore, the first
optimal divider of $T_1$ must not be at a prior position than the first optimal divider of $T_2$.

\qed
\end{proof}

\subsection{Backtracking in Optimal Seed Solver}

This section presents the pseudo code of the backtracking process in OSS.

\begin{algorithm}[t]
  \textbf{Input}: the final optimal divider of the read, $\mathsf{opt\_div}$\\
  \textbf{Output}: an array of optimal dividers of the read, $\mathsf{div\_array}$\\
  \textbf{Global data structure}: the 2-D data array $\mathsf{opt\_data[\ ][\ ]}$\\
  \textbf{Pseudocode}:\\
  \tcp{Push in the last divider}
  $\mathsf{div\_array.push(opt\_div)}$\;
  $\mathsf{prev\_div = opt\_div}$\;
  \For{$\mathit{iter} = x-1$ \KwTo $2$} {
	  $\mathsf{div} = \mathsf{opt\_data[\mathit{iter}][\mathit{prev\_div-1}]}.div$\;
	  $\mathsf{div\_array.push(opt\_div)}$\;
	  $\mathsf{prev\_div = div}$\;
  }
  \KwRet{$\mathsf{div\_array}$}\;
  \caption{Backtracking}
  \label{algo:A3}
\end{algorithm}

The pseudo code of the backtracking process is provided in Algorithm~\ref{algo:A3}. The key idea
behind the backtracking algorithm is simple: In the element of the $i^{th}$ row and the $j^{th}$
column of $opt\_data$, stores the optimal divider, $div$, of the substring $R[1...j]$. This $div$
suggests that by optimally selecting $i-1$ seeds from $R[1...div-1]$ and one seed from $R[div...j]$,
we can obtain the least frequent $i$ seeds from $R[1...j]$. From $div$ we can learn that substring
$R[div...j]$ provides the $i^{th}$ optimal seeds. Similarly, by repeating this process for the
element of $opt\_data[i-1][div-1]$, we can learn the position and length of the $(i$-$1)^{th}$
optimal seeds. We can repeat this process until we have learnt all optimal dividers of the read.

\end{document}